\pdfoutput=1
\documentclass[11pt]{article}

\usepackage{etoolbox}

\newlength{\bodyparskip}
\AtBeginDocument{%
  \setlength{\parskip}{0pt}%
}
\apptocmd{\maketitle}{%
  \setlength{\parskip}{\bodyparskip}%
}{}{}

\pretocmd{\section}{\setlength{\parskip}{0pt}}{}{}
\pretocmd{\subsection}{\setlength{\parskip}{0pt}}{}{}
\pretocmd{\subsubsection}{\setlength{\parskip}{0pt}}{}{}

\makeatletter
\let\orig@afterheading\@afterheading
\renewcommand\@afterheading{%
  \setlength{\parskip}{\bodyparskip}%
  \orig@afterheading
}
\makeatother

\usepackage[
    left=32mm,
    right=32mm,
    top=28mm,
    bottom=28mm
]{geometry}

\usepackage{amsmath}
\usepackage{amsfonts}
\usepackage{graphicx}
\usepackage{authblk}

\usepackage[hypertexnames=false]{hyperref}
\usepackage{bm}
\usepackage{newtxmath} 
\usepackage{bbm} 
\renewcommand{\mathbbm}{\vvmathbb} 
\usepackage{tensor} 

\usepackage{xcolor}

\newcommand{\vek}{\mathbf} 
\newcommand{\gec}{\bm}       

\newcommand*{\Scale}[2][4]{\scalebox{#1}{$#2$}} 

\newcommand{\rmd}{\mathrm{d}}

\newcommand{\orthvec}[1]{\hat{\mathbf{e}}_{#1}}

\title{{\Large\sffamily\bfseries The Dirac equation in arbitrary coordinate systems and curved spacetimes - An orthonormal frames approach}}

\author{Andrew DeBenedictis}
\affil{The Pacific Institute for the Mathematical Sciences\\ and\\
       Faculty of Science, Simon Fraser University,
       Burnaby, British Columbia, V5A 1S6, Canada \\ \small{(adebened@sfu.ca)}}

\author{Sa\v{s}a Iliji\'c}
\affil{University of Zagreb
       Faculty of Electrical Engineering and Computing,
       Department of Applied Physics, Unska 3, 10\,000 Zagreb, Croatia \\ \small{(sasa.ilijic@fer.hr)}}

\date{{\small{Novemver 16, 2026}}}

\begin{document}

\maketitle

\begin{abstract}
In this pedagogical article we discuss the Dirac equation in arbitrary coordinate systems. We explain the need for connections when dealing with spinor fields and present the practical steps needed to compute the Dirac operator in a general setting, allowing for arbitrary coordinate systems and position-dependent orthonormal gamma matrices, both of which can simplify computations in certain scenarios. The extension to include minimally coupled gravity is straightforward. Examples are provided, including constructing the Dirac equation in spherical and cylindrical symmetry (with or without gravity) and the Kerr spacetime. We also briefly comment on the advantages and disadvantages of various tetrads and orthonormal gamma matrices. This manuscript should be accessible to readers who have studied the Dirac equation at an introductory level (e.g., in Cartesian coordinates in flat spacetime with constant gamma matrices). The emphasis is on the use of orthonormal frames, since these correspond to the coordinate systems most familiar to readers when dealing with most field theories, and represent the physical frames where measurements are performed. The aim is to get the reader quickly to the point of being able to do calculations, so the presentation favors a concise, pragmatic approach over extended derivations. 

\end{abstract}

\maketitle


\section{Introduction}

In 1928 the now famous Dirac equation was introduced describing wave fields
corresponding to free spin one-half particles \cite{ref:dirac}. Since then the
Dirac equation has become a staple in the study of the quantum theory of fields and 
high-energy particle physics, with its natural capacity to incorporate spin and
describe antiparticles. 

When the Dirac equation is first encountered, it is usually presented under the assumption of Cartesian coordinates and constant-valued gamma matrices. Its more general form is usually reserved for more advanced research settings, such as analyses in curved spacetimes, where it is discussed mainly in what is known as the holonomic (coordinate) frame, which is not the natural frame for physical measurements, which is why we choose to concentrate on orthonormal frames in this manuscript. However, from a purely practical perspective, the equation’s role in flat spacetime is often of more immediate importance than its curved-spacetime counterpart and, even in flat spacetime, there is no single coordinate system that is superior in all settings. Different coordinate systems allow one to adapt the equation to the symmetry of the problem at hand, possibly enabling a solution in cases where another coordinate choice would obscure the path forward. Although the Dirac equation itself, being fully contracted on its spacetime indices, does not distinguish between its holonomic or orthonormal form, ideally when first encountered the general Dirac equation should be presented in orthonormal frames as this framework is actually the one that is most familiar to most people. In flat spacetime in standard Cartesian coordinates the frame is already orthonormal, so such a distinction need not be made. In more general coordinate systems (and in curved spacetimes) the distinction is manifest and the wave function, being spinorial rather than scalar or tensorial, must be handled with some care. Further to this, the gamma matrices, even in orthonormal frames in flat spacetime, may be taken to be coordinate dependent. In some scenarios, this freedom can lead to equations that are easier to solve than their constant gamma-matrix counterparts. In this manuscript we present and discuss the Dirac equation in arbitrary coordinate systems under the assumption that, even in orthonormal frames, the gamma matrices may be coordinate dependent. This discussion readily allows for the inclusion of gravity in physical frames with no modification.

Before we actually discuss the Dirac equation, let us set up the notations and conventions that will be utilized in this manuscript:
\begin{itemize}
\item There will be two sets of component indices used in this manuscript.
One set, without hats (e.g.\ $\mu$, $j$, etc.), will denote components with
respect to a \emph{coordinate basis} (or \emph{holonomic} basis).
Another set of indices, adorned with a hat (e.g.\ $\hat{\alpha}$, $\hat{a}$),
will denote components with respect to an \emph{orthonormal frame}
(generally anholonomic), to be discussed in more detail later.
\item Greek indices cover the full 4D spacetime, whereas Latin indices denote only the spatial components. For both sets of indices the Einstein summation convention is utilized throughout. 
\item The metric signature used will be the ``mostly minus'' convention (``minus 2'' since we are dealing only in four dimensions in this manuscript). That is, the matrix representing the metric has 1 positive and three negative eigenvalues. The positive one corresponds to a timelike coordinate and the three negative ones to spacelike coordinates. Further, the orthonormal metric's components are denoted by $\eta_{\hat{\alpha}\hat{\beta}}$ and the coordinate (holonomic) metric's components by $g_{\mu\nu}$.
\item The Planck constant, $\hbar$, and the speed of light, $c$ will be set equal to $1$, as will the gravitational constant $G$ where we discuss coupling to gravity. 
\end{itemize}
We shall work as much as possible in orthonormal frames, as this will be most familiar to those that have studied introductory field theories, and yield expressions most familiar from elementary calculus. It is unfortunate that the more complicated notation is used for the more utilized indices but this notation (hatted for orthonormal, unhatted for holonomic) is common in the literature where both sets of indices are present \cite{ref:MTW}, \cite{ref:schutzGR}. We briefly discuss the Dirac equation in a holonomic framework in Appendix \ref{app:holonomic}.

To clarify the meaning of the two sets of indices consider for now the infinitesimal line-element in Minkowski spacetime given in the standard spherical coordinates with minus-2 signature:
\begin{equation}
 \rmd s^{2}=g_{\mu\nu}(x)\,\rmd x^{\mu}\rmd x^{\nu} = \rmd t^{2} - \rmd r^{2}-r^{2}\,\rmd\theta^{2}-r^{2}\sin^{2}\theta\,\rmd\phi^{2}\,. \label{eq:sphpy}
\end{equation}
The \emph{holonomic metric} and the reciprocal metric can be represented by the following matrices
\begin{equation}
 \left[g_{\mu\nu}(x)\right]=\left[\begin{array}{cccc}
                                   1 & 0 & 0 & 0 \\
                                   0 & -1 & 0 & 0 \\
                                   0 & 0 & -r^{2} & 0 \\
                                   0 &  0 & 0 & -r^{2}\sin^{2}\theta
                                  \end{array}\right]\,, \qquad
\left[g^{\mu\nu}(x)\right]=\left[\begin{array}{cccc}
                                   1 & 0 & 0 & 0 \\
                                   0 & -1 & 0 & 0 \\
                                   0 & 0 & -\frac{1}{r^{2}} & 0 \\
                                   0 & 0 & 0 & -\frac{1}{r^{2}\sin^{2}\theta}
                                  \end{array}\right]\,.  \label{eq:coordmet}
\end{equation}
This metric raises and lowers the holonomic indices:
\begin{equation}
 V_{\mu}(x)=g_{\mu\nu}(x)V^{\nu}(x),\, \qquad V^{\mu}(x)=g^{\mu\nu}(x)V_{\nu}(x)\,.
\end{equation}
On the other hand, in an orthonormal frame or basis (or, loosely speaking, in ``orthonormal coordinates'') the metric takes on its familiar form
\begin{equation}
\left[\eta_{\hat{\alpha}\hat{\beta}}\right] = \left[\eta^{\hat{\alpha}\hat{\beta}}\right]= \left[\begin{array}{cccc}
                                   1 & 0 & 0 & 0 \\
                                   0 & -1 & 0 & 0 \\
                                   0 & 0 & -1 & 0 \\
                                   0 &  0 & 0 & -1
                                  \end{array}\right]\,. \label{eq:orthmet}
\end{equation}
The orthonormal basis metric raises and lowers the hatted indices:
\begin{equation}
 V_{\hat{\alpha}}(x)=\eta_{\hat{\alpha}\hat{\beta}}V^{\hat{\beta}}(x),\, \qquad V^{\hat{\alpha}}(x)=\eta^{\hat{\alpha}\hat{\beta}}V_{\hat{\beta}}(x)\,.
\end{equation}

In a holonomic basis the metric of Eq.~(\ref{eq:coordmet}) changes depending on coordinate system whereas in an orthonormal basis the metric Eq.~(\ref{eq:orthmet}) retains the same form regardless of coordinate system. (See the examples in Sec.~\ref{sec:examples}.)  This is in some sense the point of orthonormal bases. The basis vectors are, as the name implies, orthogonal and normalized to unity and the orthonormal metric, providing the inner product, reflects this with its entries being all of magnitude $1$ and on the diagonal. In standard Cartesian coordinates the holonomic basis vectors are already orthonormal, and hence there is no difference between the coordinate metric and the orthonormal metric, and this is why in most field theory and particle physics texts there is no need to introduce two sets of indices.

Now, what maps one set of indices to the other is the \emph{tetrad}, whose components are denoted as $e^{\mu}_{\;\,\hat{\alpha}}$, and the \emph{co-tetrad}, with components denoted as $e_{\mu}^{\;\,\hat{\alpha}}$. We shall often refer to both the tetrad and co-tetrad simply as the ``tetrad'', since the index placement will indicate which object is being referred to. The tetrad is of great importance when studying the Dirac equation in arbitrary coordinate systems and will be discussed in detail below. Here we simply show how the mapping is implemented. Say you have a set of vector components in a holonomic frame, $V^{\mu}$, and you wish to know the components of this vector in an orthonormal basis. One simply multiplies $V^{\mu}$ with the tetrad as
\begin{equation}
 V^{\hat{\alpha}}(x)=e_{\mu}^{\;\,\hat{\alpha}}(x)\,V^{\mu}(x)\,, \qquad \text{and} \qquad V^{\mu}(x)=e^{\mu}_{\;\,\hat{\alpha}}(x)\,V^{\hat{\alpha}}(x)\,. \label{eq:indexswitch}
\end{equation}
A nice
way to think of Eq.~(\ref{eq:indexswitch}) is: $V^{\mu}(x)$ are the components you
write down using the coordinate grid (which may be neither orthogonal nor of
unit length), while $V^{\hat{\alpha}}(x)$ are the components a local observer would
report after projecting onto their orthonormal axes (their set of orthogonal
and normalized rulers and clocks).
So, for the case of derivatives we have that
\begin{equation}
 \partial_{\hat{\alpha}}:=e^{\mu}_{\;\,\hat{\alpha}}(x)\,\partial_{\mu}\,, \label{eq:orthpartial}
\end{equation}
which yields the appropriate normalization factors for orthonormal derivatives in curvilinear coordinates (e.g., in the usual spherical coordinates,  $\partial_{\hat{\phi}}=\partial_{\phi}/(r\sin\theta)$ with $\partial _{\phi}:=\partial/\partial\phi$ etc.) 

Finally, it will be assumed that there is no spacetime torsion. 

Now we are ready to discuss the Dirac equation. When first encountered, the Dirac equation is usually presented in this form:
\begin{equation}
 \Bigl[i\gamma^{\hat{\alpha}}\partial_{\hat{\alpha}}  -m\Bigr]\Psi(x)  = 0\,, \label{eq:diraceqn}
\end{equation}
where the matrices $\gamma^{\hat{\alpha}}$  are the Dirac gamma matrices, usually presented in the Pauli-Dirac or Weyl (chiral) representations where their components are constant. The gamma matrices provide a matrix representation of the metric, $\eta^{\hat{\alpha}\hat{\beta}}$ via the following \emph{Clifford-algebra relation}
\begin{equation}
 \left\{\gamma^{\hat{\alpha}},\, \gamma^{\hat{\beta}}\right\}:=\gamma^{\hat{\alpha}}\gamma^{\hat{\beta}}+\gamma^{\hat{\beta}}\gamma^{\hat{\alpha}}=2\eta^{\hat{\alpha}\hat{\beta}} \mathbbm{1}\,. \label{eq:clifalg}
\end{equation}

When presented in the form Eq.~(\ref{eq:diraceqn}), it is being assumed that the coordinate system is a Cartesian coordinate system, and that the $\gamma$-matrices are constant. The aim here is to discuss how Eq.~(\ref{eq:diraceqn}) must be generalized in order to accommodate arbitrary coordinate systems and non-constant $\gamma$-matrices while still retaining the orthonormal (physical) frame structure of Eq.~(\ref{eq:diraceqn}). This formalism will also open up the avenue to coupling the Dirac equation to gravity, which follows in a straightforward manner once arbitrary coordinates are accommodated.

Before continuing on this task we briefly introduce the Dirac adjoint, here given by
\begin{equation}
\overline{\Psi}(x):=\Psi^\dagger(x)\,\gamma^{\hat{0}}(x)\,, \label{eq:psibar}
\end{equation}
and assume that the
orthonormal Dirac matrices are chosen to satisfy the
following hermiticity condition:
\begin{equation}
\big(\gamma^{\hat{\alpha}}(x)\big)^{\dagger}
=\gamma^{\hat{0}}(x)\,\gamma^{\hat{\alpha}}(x)\,\gamma^{\hat{0}}(x)\, ,
\label{eq:orthgamherm}
\end{equation}
where we have relaxed the constancy condition on the gamma matrices.  We should note here that although we write $\gamma^{\hat\alpha}(x)$, we do \emph{not}
allow a completely arbitrary set of spacetime dependent $\gamma$ matrices that satisfy
\eqref{eq:clifalg}. Rather, we restrict attention to choices for which
$\gamma^{\hat0}(x)$ hermitizes the Clifford representation, i.e.\ the gamma matrices are
assumed to satisfy the pointwise hermiticity condition \eqref{eq:orthgamherm}. More general $x$-dependent Clifford representations (not generated by local spin-frame transformations compatible with \eqref{eq:orthgamherm}  on standard constant orthonormal gamma matrices)
would require additional structure, which we do not introduce here. 
Together with the Clifford algebra \eqref{eq:clifalg}, which holds even for non-constant gammas, the condition Eq.~(\ref{eq:orthgamherm}) implies
$\big(\gamma^{\hat{0}}(x)\big)^{\dagger}=\gamma^{\hat{0}}(x)$ and
$\big(\gamma^{\hat{b}}(x)\big)^{\dagger}=-\gamma^{\hat{b}}(x)$
for spatial indices $\hat{b}=1,2,3$.

\section{Tetrads and frames}
\begin{figure}
\begin{center}
\includegraphics[width=7.50cm]{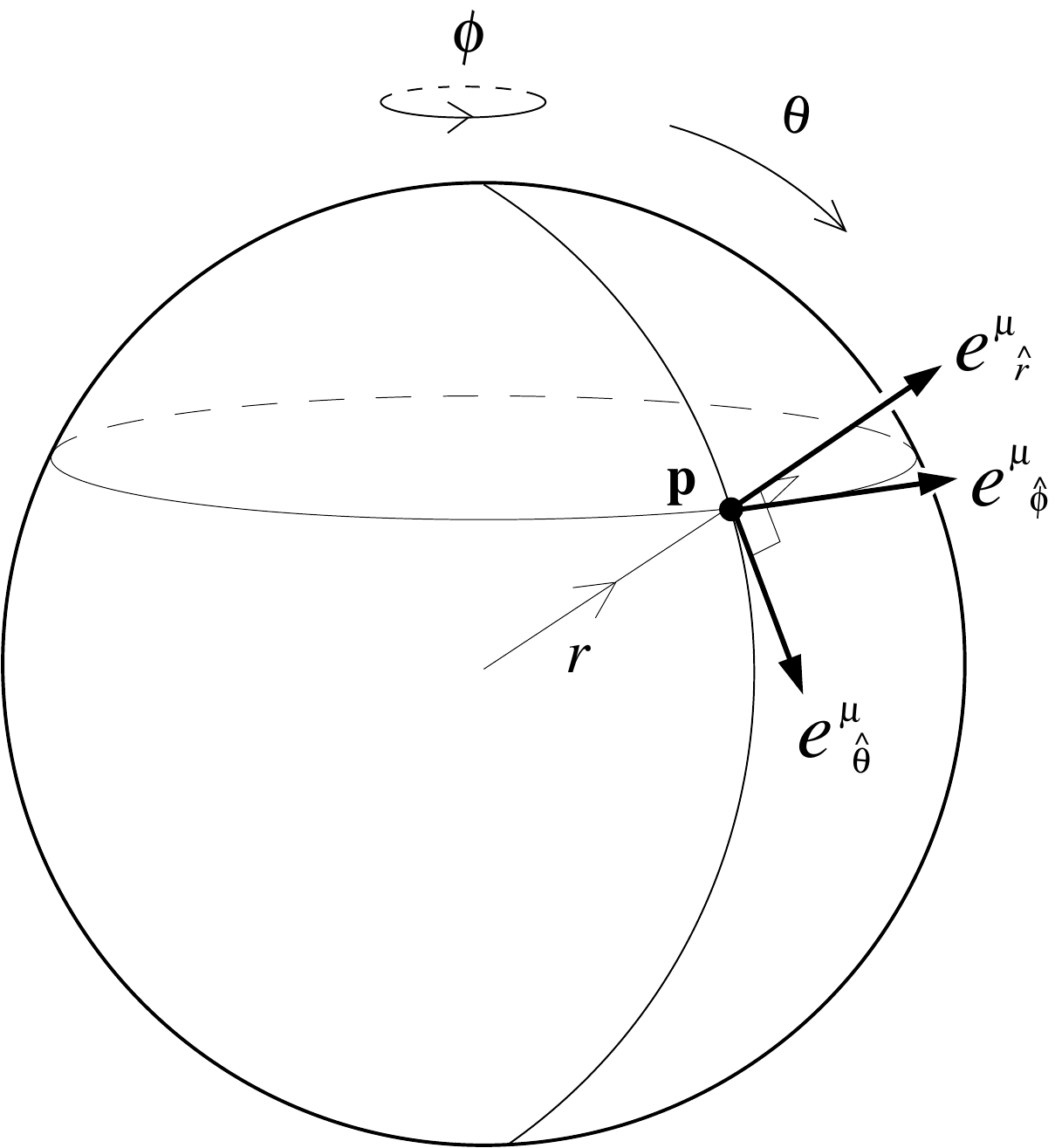}
\end{center} \caption{\label{fig:tet}\small{
A set of three vectors making up an orthonormal tetrad at the point $\vek{p}$
(time direction suppressed). In this particular example, the tetrad vectors
point in the direction of increasing spherical coordinate values. The full
tetrad would also include a vector pointing in the time direction.}}
\end{figure}

Here we will more properly discuss the (co)tetrad (or (co)vierbein, as it is sometimes called), which was briefly introduced in the Introduction. This is a set of four orthonormal (co)vectors
defined at every point in spacetime. The orthonormal index on $e^{\mu}_{\;\,\hat{\alpha}}$ or $e_{\mu}^{\;\,\hat{\alpha}}$ denotes which
orthonormal (co)vector we are referring to, and the coordinate index tells us which
component of this vector is being referred to. An example of an orthonormal
tetrad in spherical coordinates (with the time direction suppressed) is
illustrated in Fig.~\ref{fig:tet}.

The tetrad provides you with a local frame, that is, a set of orthonormal ``rulers and
clocks'' attached to each point of spacetime. The basic reason we introduce it
is that orthonormal coordinates are particularly useful, and easy to visualize.
The coordinates $x^\mu$ are mainly a way to \emph{label} events, and the
coordinate basis vectors $\partial_\mu$ that come with those labels are not
usually unit-length nor necessarily mutually orthogonal. (Spherical coordinates
are the everyday example: the coordinate directions are convenient, and
orthogonal, but they are not normalized to unity so do not form an
ortho\emph{normal} set.) The tetrad is the relativistic version of switching to
an orthonormal basis: at each point it provides one timelike direction and
three spacelike directions, all mutually orthogonal, that behave like an
ordinary special-relativistic inertial frame. (We do not consider tetrads with null vectors here.)

As discussed above, we will use two kinds of indices. Unhatted indices
$\mu,\nu,\dots$ refer to components in the coordinate basis, while hatted
indices $\hat{\alpha},\hat{\beta},\dots$ refer to components in the local
orthonormal frame. The tetrad field is the
translator between the two as in Eq.~(\ref{eq:indexswitch}). 

What determines orthonormality is the spacetime metric, meaning
\begin{equation}
g^{\mu\nu}(x)\,e_{\mu}^{\;\,\hat{\alpha}}(x)\,e_{\nu}{}^{\hat{\beta}}(x)
= \eta^{\hat{\alpha}\hat{\beta}}.
\label{eq:tetradorthonormal}
\end{equation}
Similarly, the spacetime holonomic metric can be written as
\begin{equation}
g_{\mu\nu}(x)
= \eta_{\hat{\alpha}\hat{\beta}}\,
  e_{\mu}^{\;\,\hat{\alpha}}(x)\,e_{\nu}{}^{\hat{\beta}}(x)\,.
\label{eq:metricfromtetrad}
\end{equation}
So even if the coordinate description looks ``non-orthonormal,'' the tetrad
frame always looks locally Minkowskian, as expressed by
Eq.~(\ref{eq:tetradorthonormal}).

We briefly outline here the method to calculate the tetrad vectors.
A convenient way to think about the tetrad is that it is a kind of ``square
root'' of the metric, via Eq.~(\ref{eq:metricfromtetrad}). By definition, a tetrad field $e_{\mu}{}^{\hat{\alpha}}(x)$
is any set of four linearly independent vectors (equivalently, a coframe of four
$1$-forms) that reproduces the spacetime metric via Eq.~(\ref{eq:metricfromtetrad}).
In matrix language this is $g=e^{T}\eta e$. Thus, at each point, choosing a
tetrad is nothing more than choosing an orthonormal basis for the tangent space
with respect to $g_{\mu\nu}(x)$.
In Appendix \ref{app:tetrad} we illustrate one method that can be used to explicitly compute a tetrad.

Any two such tetrad sets are related by a local Lorentz transformation; this
is simply the built-in freedom to rotate or boost the local orthonormal axes
at each point without changing the metric. Let
$\Lambda^{\hat{\alpha}'}{}_{\hat{\beta}}(x)$ denote a possibly
position-dependent Lorentz transformation taking the old local frame to the
new one. Then
\begin{equation}
e_{\mu}^{\;\,\hat{\alpha}}(x)
\ \longrightarrow\
e_{\mu}^{\;\,\hat{\alpha}'}(x)
= \Lambda^{\hat{\alpha}'}{}_{\hat{\beta}}(x)\,e_{\mu}^{\;\,\hat{\beta}}(x).
\label{eq:tetradLorentz}
\end{equation}
Under this change of local frame, hatted components transform exactly as in
special relativity; for example,
\begin{equation}
V^{\hat{\alpha}}(x)
\ \longrightarrow\
V^{\hat{\alpha}'}(x)
= \Lambda^{\hat{\alpha}'}{}_{\hat{\beta}}(x)\,V^{\hat{\beta}}(x)\,.
\label{eq:VlocalLorentz}
\end{equation}
In other words, Lorentz transformations ``act on hatted
indices'': they are changes of the \emph{local orthonormal frame} as in
Eq.~(\ref{eq:tetradLorentz}), not changes of coordinates.


Now let us move to spinors. A Dirac spinor is not a spacetime
vector. Under a change of the local orthonormal frame, the hatted components
of a vector transform with the Lorentz matrix $\Lambda$, as in
Eq.~(\ref{eq:VlocalLorentz}), whereas a Dirac spinor transforms according to
the corresponding spinor representation $S(\Lambda)$. The tetrad is what
supplies the local Minkowski frame needed to even \emph{state} this
transformation law at each point. So when we change the tetrad according to
$\Lambda^{\hat{\alpha}'}{}_{\hat{\beta}}(x)$, as in
Eq.~(\ref{eq:tetradLorentz}), the spinor field transforms by the usual spinor
representation associated with the same Lorentz transformation:
\begin{equation}
\Psi(x)\ \longrightarrow\ \Psi'(x)
= S\!\left(\Lambda(x)\right)\,\Psi(x),
\label{eq:spinorlocalLorentz}
\end{equation}
where $\Lambda(x)$ is the Lorentz matrix with components
$\Lambda^{\hat{\alpha}'}{}_{\hat{\beta}}(x)$, and $S(\Lambda(x))$ is the
corresponding $4\times 4$ matrix acting on the bispinor components.

At this stage the plan is as follows: we want the Dirac equation to look
locally like its familiar special-relativistic form when expressed in the
orthonormal frame, because that is the frame where spinors ``live.'' The tetrad
provides that frame and the right notion of local Lorentz transformations,
summarized in Eq.~(\ref{eq:tetradLorentz}) and Eq.~(\ref{eq:spinorlocalLorentz}). The
remaining piece---how to take derivatives of a spinor when the local frame
varies from point to point---is what we will introduce near the end of the following section. This will then allow us to write the Dirac equation in a coordinate independent way.


\section{The need for connections}

In order to motivate the need for a connection, we will in this section first study something that is seemingly unrelated to the Dirac equation and spinors. Let us consider an ordinary vector $\vek{V}$ in $\mathbb{R}^{3}$ expressed in the orthonormal basis:
\begin{equation}
 \vek{V}=V^{\hat{b}}(x)\,\orthvec{\hat{b}}\,, \label{eq:vector}
\end{equation}
where $V^{\hat{b}}(x)$ are the components of the vector field $\mathbf{V}(x)$ and $\orthvec{\hat{b}}$ are a set of three orthonormal vectors making up the basis (i.e. here the index on the basis vector is telling us which vector, not the components of the vector). For example, in Cartesian coordinates we could have
\begin{equation}
 \orthvec{\hat{1}}=\hat{\vek{x}}=\left[\begin{array}{c}
                                  1 \\
                                  0\\
                                  0
                                 \end{array}\right]\,, \quad 
\orthvec{\hat{2}}=\hat{\vek{y}}=\left[\begin{array}{c}
                                  0 \\
                                  1\\
                                  0
                                 \end{array}\right]\,, \quad 
\orthvec{\hat{3}}=\hat{\vek{z}}=\left[\begin{array}{c}
                                  0 \\
                                  0\\
                                  1
                                 \end{array}\right]\,, \nonumber 
\end{equation}
and, important for this discussion, their directions are independent of their location.
In spherical coordinates the situation is similar. However, there is an important difference and that is that in spherical coordinates the $\orthvec{\hat{i}}$ \emph{do} change direction depending on their locations. Think of, for example, of the vector $\hat{\gec{\upphi}}$ which points around the $z$-axis and so its direction depends on the location around the $z$-axis.

Now let us take the derivative  of the vector $\vek{V}$ in Eq.~(\ref{eq:orthpartial}) along the orthonormal direction $\hat{i}$ 
\begin{equation}
 \frac{\partial \vek{V}}{\partial x^{\hat{i}}} = \frac{\partial V^{\hat{b}}(x)}{\partial x^{\hat{i}}} \orthvec{\hat{b}} + V^{\hat{b}}(x)\,\frac{\partial \orthvec{\hat{b}}}{\partial x^{\hat{i}}}\,. \label{eq:vecderiv}
\end{equation}
Let's specifically study the last term in the above expression, in particular the part  $\frac{\partial \orthvec{\hat{b}}}{\partial x^{\hat{i}}}$. Now, whatever this is, for \emph{each individual value} of the index $\hat{i}$ and $\hat{b}$ it is a vector. Therefore, whatever this object is, for any fixed values of  $\hat{i}$ and $\hat{b}$  it can be expressed as a linear combination of the basis vectors. That is,
\begin{equation}
 \frac{\partial \orthvec{\hat{b}}}{\partial x^{\hat{i}}} =: {\omega_{\hat{i}}^{\;\,\hat{\ell}}}_{\hat{b}}(x)\, \orthvec{\hat{\ell}}\,. \label{eq:basisderiv}
\end{equation}
The components ${\omega_{\hat{i}}^{\;\,\hat{\ell}}}_{\hat{b}}(x)$ are called the
\emph{fully orthonormal Ricci rotation coefficients}\footnote{It should be noted
that there is no universal agreement on the index placement so the reader should
check the convention in the literature they are reading.} The 
Ricci rotation coefficients represent
the $\hat{\ell}^{\text{th}}$ component of the derivative of the basis vector
$\orthvec{\hat{b}}$ with respect to the coordinate $x^{\hat{i}}$, i.e. it is the
$\hat{\ell}^{\text{th}}$ component of $\partial_{\hat{i}} \orthvec{\hat{b}}$. Notice now that the derivative of Eq.~(\ref{eq:vecderiv}) can be written, after relabeling some dummy indices, as
\begin{equation}
\frac{\partial \vek{V}}{\partial x^{\hat{i}}}
= \left[\partial_{\hat{i}} V^{\hat{b}}(x) + V^{\hat{k}}(x)\,{\omega_{\hat{i}}^{\;\,\hat{b}}}_{\hat{k}}(x)\right]\orthvec{\hat{b}} \,.
\label{eq:covdif}
\end{equation}
The quantity in square brackets is known as the components of the \emph{orthonormal covariant derivative} of the vector $\vek{V}$. As can be seen, it contains the ordinary (orthonormal) partial derivative plus a term that encodes the change of basis vectors under derivation. The  ${\omega_{\hat{i}}^{\;\,\hat{b}}}_{\hat{k}}(x)$ make up the components of a \emph{connection}. Connections are needed in derivatives whenever there is a non-constant basis involved. They are called ``rotation'' coefficients since, due to the unit normalization, the basis vectors can only rotate, but not change length, when their derivatives are taken. Also, notice that since in Cartesian components the basis vectors are all constant, their derivatives are zero, and hence the Ricci rotation coefficients in Cartesian coordinates are zero via Eq.~(\ref{eq:basisderiv}).

Let us now generalize the above to \emph{relativistic} vectors. Note that here, dealing with a relativistic theory, it is important if an index is written as contravariant (superscript) or covariant (subscript). Writing a vector
\begin{equation}
\vek{V} = V^{\hat{\alpha}}(x)\,\orthvec{\hat{\alpha}}\,,
\nonumber
\end{equation}
we find
\begin{equation}
\partial_{\hat{\beta}} \vek{V}
= \left(\partial_{\hat{\beta}} V^{\hat{\alpha}}(x)\right)\,\orthvec{\hat{\alpha}}
+ V^{\hat{\alpha}}(x)\,(\partial_{\hat{\beta}} \orthvec{\hat{\alpha}})\,,
\nonumber
\end{equation}
where $\partial_{\hat{\beta}}:= e^{\nu}_{\;\,\hat{\beta}}(x)\partial_{\nu}$ denotes the
directional derivative along the orthonormal frame in spacetime. The second term is again what
forces us to introduce the  Ricci rotation coefficients via
\begin{equation}
\partial_{\hat{\beta}} \orthvec{\hat{\alpha}}
= \tensor{\omega}{_{\hat{\beta}}^{\hat{\gamma}}_{\!\hat{\alpha}}}(x)\,\orthvec{\hat{\gamma}} \,.
\nonumber
\end{equation}
In words: the orthonormal basis vectors ``rotate'' from point to point, and the
$\tensor{\omega}{_{\hat{\beta}}^{\hat{\gamma}}_{\!\hat{\alpha}}}\orthvec{\hat{\gamma}}(x)$ precisely measure that rotation,
allowing us to define a covariant derivative of the vector components.

Since the aim is to get the reader to a stage where they can perform
calculations quickly, we will take a rather brief and pragmatic approach
here, without detailed derivations. We need two things for this: one is
the previously discussed tetrad and the other is an explicit expression
for our Ricci rotation coefficients, which is furnished by
\begin{equation}
\omega_{\hat{\delta}}^{\;\,\hat{\alpha}\hat{\beta}}
=
e^{\mu}_{\;\,\hat{\delta}}\left\{
e^{\nu\hat{\alpha}} \partial_{[\mu} e_{\nu]}{}^{\hat{\beta}}
- e^{\nu\hat{\beta}}\partial_{[\mu}e_{\nu]}{}^{\hat{\alpha}}
- e^{\rho\hat{\alpha}} e^{\sigma\hat{\beta}}
\partial_{[\rho}e_{\sigma] \hat{\gamma}}\,
e_{\mu}^{\;\,\hat{\gamma}}
\right\}
=: e^{\mu}_{\;\,\hat{\delta}}\,\omega_{\mu}^{\;\,\hat{\alpha}\hat{\beta}}\,.
\label{eq:rrcformula}
\end{equation}
Here the square brackets around indices denote antisymmetrization on
those indices: $ X_{[\mu\nu]}:=\left(X_{\mu\nu}-X_{\nu\mu}\right)/2$.

The formula in Eq.~(\ref{eq:rrcformula}) is such that the Ricci rotation
coefficients have the following property:
\begin{equation}
\omega_{\hat{\delta}}^{\;\,\hat{\alpha}\hat{\beta}}
=-\omega_{\hat{\delta}}^{\;\,\hat{\beta}\hat{\alpha}}\,,
\label{eq:RRCantisym}
\end{equation}
i.e.\ they are antisymmetric under exchange of the last two indices.
(The object $\omega_{\mu}^{\;\,\hat{\alpha}\hat{\beta}}$ introduced on
the far right of Eq.~(\ref{eq:rrcformula}) is known as the
\emph{spin connection}. We shall not need it here, but it is often used
in the study of spinors when one does not work entirely in orthonormal
frames (see Eq.~(\ref{eq:spinaffinecon}).)

The expression of Eq.~(\ref{eq:rrcformula}) is fixed by the requirement that
our orthonormal frame can be differentiated in a way that respects the
local Lorentz geometry. Concretely, the Ricci rotation coefficients are
chosen so that the coframe fields $e_{\mu}{}^{\hat{\alpha}}$ satisfy
what is known as the torsion-free condition, which may be written in the
purely frame-based form 
\begin{equation}
e^{\mu}{}_{\hat{\gamma}}\,e^{\nu}{}_{\hat{\delta}}\,
\partial_{[\mu} e_{\nu]}{}^{\hat{\alpha}}
+\frac{1}{2}\Big(
\omega_{\hat{\gamma}}{}^{\hat{\alpha}}{}_{\hat{\delta}}
-\omega_{\hat{\delta}}{}^{\hat{\alpha}}{}_{\hat{\gamma}}
\Big)
=0 \, .
\label{eq:tetrad_covconst_frame_full}
\end{equation}
The reason for this requirement is that it guarantees the orthonormal
frame remains orthonormal when we compare nearby points. Equivalently, mathematically,
it picks out the standard torsion-free, metric-compatible way of
differentiating objects with hatted indices, and thereby determines the
Ricci rotation coefficients entirely in terms of the tetrad and its
first derivatives. This is the geometric input needed later when we
couple spinors to curvilinear coordinates or to nontrivial frames.

Further, we require the partial derivative in spacetime in the orthonormal basis (i.e. the partial derivative in the $\hat{\alpha}$ orthonormal direction), which is provided by Eq.~(\ref{eq:orthpartial}).

Now, since in general we are not restricted to using Cartesian
coordinates, if we choose curvilinear coordinates the coordinate basis
vectors $\partial_\mu$ are generally not orthonormal, so they are not the
natural directions in which the Dirac equation is usually written. Nor
are they generally normalized to unity. The way we recover the familiar
``Cartesian'' viewpoint locally is to introduce an orthonormal frame at
each point. This is the role of the tetrad $e_{\mu}{}^{\hat{\alpha}}(x)$,
which lets us work with orthonormal components even though the underlying
coordinates are arbitrary. In particular, the tetrad lets us define
directional derivatives along the orthonormal axes by
Eq.~(\ref{eq:orthpartial}), so that we can speak of ``time- and
space-derivatives measured by a local inertial observer,'' even though we
are differentiating with respect to the underlying coordinates $x^\mu$.

The price we pay for this local orthonormal description is that, as  before with $\mathbb{R}^{3}$, the
orthonormal axes can vary from point to point. This introduces a freedom
absent in global Cartesian coordinates: at each point we may rotate or
boost the orthonormal frame without changing any physics. Concretely, we
may replace one tetrad by another related by a local Lorentz
transformation $\Lambda(x)$. Ordinary vectors written in orthonormal
components transform in the expected way,
\begin{equation}
V^{\hat{\alpha}}(x)\;\to\;V^{\hat{\alpha}'}(x)
=\Lambda^{\hat{\alpha}'}{}_{\hat{\beta}}(x)\,V^{\hat{\beta}}(x),
\end{equation}
so the hatted indices behave like special-relativistic indices.

This is why derivatives needed care, as we saw in $\mathbb{R}^{3}$. Even if we never perform any explicit
Lorentz transformation, the orthonormal basis itself is generally
position-dependent, so a derivative differentiates the changing basis as well as the
components. However, here with relativity, we can use an equivalent statement, but now in terms of Lorentz transformations: if we re-choose the tetrad by a
position-dependent Lorentz transformation $\Lambda(x)$, the same problem
appears: $\partial_{\hat{\alpha}}V^{\hat{\beta}}$ does not transform
cleanly. The covariant derivative corrects for this. In a fully
orthonormal frame it is written as
\begin{equation}
\nabla_{\hat{\alpha}}V^{\hat{\beta}}(x)
=
\partial_{\hat{\alpha}}V^{\hat{\beta}}(x)
+
\omega_{\hat{\alpha}}{}^{\hat{\beta}}{}_{\hat{\gamma}}(x)\,
V^{\hat{\gamma}}(x)\,,
\end{equation}
where $\omega_{\hat{\alpha}}{}^{\hat{\beta}}{}_{\hat{\gamma}}$ (the Ricci
rotation coefficients), as before, encodes how the orthonormal basis
varies from point to point. Equivalently, it encodes the infinitesimal
rotation/boost relating neighboring local frames.


\subsection{Connections in the Dirac equation}

We now graduate to the Dirac equation.  In order to motivate the requirement for a connection here let us first discuss briefly the Dirac equation in Cartesian coordinates. For this part of the discussion we will take as the orthonormal gamma matrices the following standard Pauli-Dirac representation
\begin{equation}
\gamma^{\hat{0}}=\left[\begin{array}{cc}
                        \mathbbm{1} & \mathbbm{0} \\
                        \mathbbm{0} & -\mathbbm{1}
                       \end{array}\right]\,, \quad
\gamma^{\hat{b}}=\left[\begin{array}{cc}
                        \mathbbm{0} & \sigma^{b} \\
                        -\sigma^{b} & \mathbbm{0}
                       \end{array}\right]\,,  \label{eq:paulidiracgams1}
\end{equation}
with the $\sigma^{b}$ the standard Pauli matrices here yielding
\begin{equation}
\gamma^{\hat 0}=
\begin{bmatrix}
1&0&0&0\\
0&1&0&0\\
0&0&-1&0\\
0&0&0&-1
\end{bmatrix}\,,
\quad
\gamma^{\hat 1}=
\begin{bmatrix}
0&0&0&1\\
0&0&1&0\\
0&-1&0&0\\
-1&0&0&0
\end{bmatrix}\,,
\quad
\gamma^{\hat 2}=
\begin{bmatrix}
0&0&0&-i\\
0&0&i&0\\
0&i&0&0\\
-i&0&0&0
\end{bmatrix}\,,
\quad
\gamma^{\hat 3}=
\begin{bmatrix}
0&0&1&0\\
0&0&0&-1\\
-1&0&0&0\\
0&1&0&0
\end{bmatrix}\,. \label{eq:paulidiracgams2}
\end{equation}
It can easily be checked that these matrices obey the Clifford algebra relation in Eq.~(\ref{eq:clifalg}). 

Let us also for the moment restrict ourselves to a frame where the particle is at rest. This means that the spatial momentum  of the particle is zero, which in the language of quantum mechanics implies that the spinor components are independent of position ($\widehat{\vek{p}}\,\Psi(x)=0$; in components this means
$\widehat{p}^{\,\hat{b}}\Psi(x)
= i\,\partial^{\hat{b}}\Psi(x)
= -\,i\,\partial_{\hat{b}}\Psi(x)
= 0$.)
 Under these assumptions the Dirac equation, Eq.~(\ref{eq:diraceqn}), takes on the form
\begin{equation}
 i\gamma^{\hat{0}}\partial_{\hat{0}}\Psi(t)-m\Psi(t)=0\,. \label{eq:posindepdirac}
\end{equation}
The solutions to the above equation are well known
(see, for example, \cite{ref:griffithsparticles},
which also provides a nice discussion of their properties)
so we just quote them here.
One set of four unnormalized solutions is furnished by
\begin{equation}
 \Psi(x)_{1}=e^{-imt}\left[\begin{array}{c}
                         1\\0\\0\\0
                        \end{array}\right]\,,\quad 
\Psi(x)_{2}=e^{-imt}\left[\begin{array}{c}
                         0\\1\\0\\0
                        \end{array}\right]\,,\quad
\Psi(x)_{3}=e^{imt}\left[\begin{array}{c}
                         0\\0\\1\\0
                        \end{array}\right]\,,\quad 
\Psi(x)_{4}=e^{imt}\left[\begin{array}{c}
                         0\\0\\0\\1
                        \end{array}\right]\,. \label{eq:staticsols}
\end{equation}
These solutions describe the following physical scenarios \cite{ref:griffithsparticles}:
\begin{align}
\Psi(x)_{1}&=\text{particle with spin up}, \nonumber \\
\Psi(x)_{2}&=\text{particle with spin down}, \nonumber \\
\Psi(x)_{3}&=\text{anti-particle with spin up}, \nonumber \\
\Psi(x)_{4}&=\text{anti-particle with spin down}. \nonumber 
\end{align}
Note that the four column vectors in Eq.~(\ref{eq:staticsols}) form a convenient basis of
\emph{rest-frame energy eigenspinors} (in the chosen Dirac representation and for
a chosen spin quantization axis). In this sense they play a role analogous to the previously discussed 
orthonormal basis $\{\orthvec{\hat{\beta}}\}$, except that the
``directions'' now refer to the internal spinor space: particle/antiparticle
(positive/negative frequency) and spin up/down. Any rest-frame spinor can be
expanded as a linear combination of these basis spinors. 
Put in loose language, these four basis spinors define four independent
``directions'' (axes) in the internal spinor space, corresponding in
this basis to particle/antiparticle and spin up/down.\footnote{In a quantum-field
theory setting, the negative-frequency modes are reinterpreted as antiparticle
degrees of freedom.} Each basis spinor may be thought of as ``pointing'' along
one such axis. 

Because of the presence of a basis, the Dirac spinor in arbitrary coordinate
systems \emph{conceptually} behaves in a similar manner to the vectors
that we discussed previously when taking derivatives. In the earlier vector discussion
the key point was fairly simple: if a vector field is expanded in a
position-dependent basis (e.g. a spherical \emph{orthonormal} basis where the
direction that the basis vectors point in depends on their position in space),
then taking a derivative necessarily differentiates both the components of the
vector and the basis itself. In the relativistic case local Lorentz transformations provide an equivalent picture.

For Dirac spinors the object that
``rotates'' is now the \emph{spinor basis} (or equivalently, the local choice of
spin axes). A spinor field $\Psi(x)$ has four complex components, but the physical
meaning of these components (for example, what we call ``spin up'' and ``spin
down'') is tied to a choice of local orthonormal frame (i.e. the tetrad we choose,
not the coordinate basis). In Cartesian coordinates one can choose this frame
globally so that it does not vary with position, and then a plain partial
derivative of the spinor components is sufficient. In a curvilinear coordinate system,
however, the natural orthonormal frame is generally position-dependent (as in
spherical coordinates), and as that frame rotates from point to point the
associated spinor basis rotates with it. Therefore, just as for vectors, the
derivative of the spinor $\Psi(x)$ must be corrected by an additional term (a connection!)
that accounts for the rotation of the chosen local frame.

This correction is implemented by introducing the components of the
\emph{spinor covariant derivative} in an analogous way we did for ordinary vectors:
\begin{equation}
\nabla_{\hat{\alpha}} \Psi(x) := \partial_{\hat{\alpha}} \Psi(x) + \Gamma_{\hat{\alpha}}(x)\,\Psi(x)\,,
\label{eq:spincovdif}
\end{equation}
where $\Gamma_{\hat{\alpha}}(x)$ are known as the \emph{Fock-Ivanenko coefficients}. For each value of $\hat{\alpha}$ these are matrices multiplying $\Psi(x)$. They play a somewhat analogous role for spinors that the Ricci rotation coefficients play for vectors:
here they compensate for the position dependence (rotation) of the local orthonormal
frame, ensuring that the Dirac equation has the same form in any coordinate
system. Note the similarity between this formula and the expression in brackets in Eq.~(\ref{eq:covdif}). We shall discuss the properties of $\Gamma_{\hat{\alpha}}(x)$, as well as how to compute it, in the next section.

We would also like to know what the covariant derivative on  $\overline{\Psi}(x)$ looks like. For this, consider the object $\overline{\Psi}(x)\Psi(x)$. Since $\overline{\Psi}(x)\Psi(x)$ is a scalar quantity, the following should hold:
\begin{equation}
 \nabla_{\hat{\alpha}}\left(\overline{\Psi}(x)\Psi(x)\right)=\partial_{\hat{\alpha}}\left(\overline{\Psi}(x)\Psi(x)\right)\,, \label{eq:bigtest}
\end{equation}
that is, the quantity:
\begin{equation}
 \left(\nabla_{\hat{\alpha}}\overline{\Psi}(x)\right)\Psi(x) + \overline{\Psi}(x)\,\nabla_{\hat{\alpha}}\Psi(x) = \left(\partial_{\hat{\alpha}}\overline{\Psi}(x)\right)\Psi(x) + \overline{\Psi}(x)\,\partial_{\hat{\alpha}}\Psi(x)\,. \nonumber
\end{equation}
For this to hold with Eq.~(\ref{eq:spincovdif}) we require that
\begin{equation}
\nabla_{\hat{\alpha}}\overline{\Psi}(x) := \partial_{\hat{\alpha}}\overline{\Psi}(x) -
\overline{\Psi}(x)\,\Gamma_{\hat{\alpha}}(x)\,. \label{eq:psibarcovdif}
\end{equation}

It should be emphasized that the purpose of $\Gamma_{\hat{\alpha}}(x)$ is not
to modify the particle/antiparticle interpretation. Rather, $\Gamma_{\hat{\alpha}}(x)$
is required because we have chosen a position-dependent local orthonormal frame:
it compensates for the local Lorentz rotation/boost of the tetrad from point to
point, so that derivatives of spinor components are taken in a frame-covariant
way. For inertial frames related by proper (orthochronous) Lorentz
transformations, the particle/antiparticle (positive/negative energy) split is
preserved; it is tied to the chosen time direction and the corresponding notion
of energy, not to the introduction of the spin connection.

Using the prescription in Eq.~(\ref{eq:spincovdif}), the Dirac equation in arbitrary coordinate systems is written simply by replacing the partial derivative in Eq.~(\ref{eq:diraceqn}) with the covariant derivative of Eq.~(\ref{eq:spincovdif}):
\begin{equation}
\boxed{ \left(i\gamma^{\hat{\alpha}}\nabla_{\hat{\alpha}}-m\right)\Psi(x) = \Bigl[i\gamma^{\hat{\alpha}}\left(\partial_{\hat{\alpha}} + \Gamma_{\hat{\alpha}}(x)\right) -m\Bigr]\Psi(x)  = 0}\,. \label{eq:arbdireqn}
\end{equation}
So, if we know how to compute the $\Gamma_{\hat{\hat{\alpha}}}(x)$, we know the Dirac equation in our chosen coordinate system.
We will discuss the properties of the Fock-Ivanenko connection in what follows.


\section{Properties of the connection and spinor covariant derivative}\label{sec:covdif}
A Dirac spinor $\Psi(x)$ is not a Lorentz vector; it does
not carry a Lorentz index $\hat{\alpha}$. Nevertheless, it still transforms
under Lorentz transformations, but in a different way. Under a local Lorentz
transformation, a spinor transforms as
\begin{equation}
\Psi(x)\;\to\;\Psi'(x)=S(\Lambda(x))\,\Psi(x),
\end{equation}
where $S(\Lambda)$ is the $4\times 4$ spinor matrix associated with the same
Lorentz transformation. (This is the first place where spinors differ in an
essential way from vectors: the matrix acting on the spinor is not $\Lambda$
itself, but another matrix built from it.)

If we try to differentiate $\Psi(x)$ with an ordinary derivative, the same
``moving basis'' problem appears, but now in spinor space:
\begin{equation}
\partial_{\hat{\alpha}}\Psi(x)'
=
\partial_{\hat{\alpha}}\!\big(S(x)\Psi(x)\big)
=
S(x)\,\partial_{\hat{\alpha}}\Psi(x)
+
(\partial_{\hat{\alpha}}S(x))\,\Psi(x).
\end{equation}
The extra term $(\partial_{\hat{\alpha}}S(x))\Psi(x)$ means that
$\partial_{\hat{\alpha}}\Psi(x)$ does not transform in the same way as $\Psi(x)$.
Therefore $\partial_{\hat{\alpha}}\Psi(x)$ is not a suitable object to place into a
Dirac equation if we want the equation to be independent of our arbitrary
choice of local orthonormal frame.

The cure is the same in spirit as for vectors: we must add a connection term
that compensates for the position-dependence of the local frame. For spinors,
this takes the form of a matrix-valued connection,
\begin{equation}
\nabla_{\hat{\alpha}}\Psi(x)
:=
\partial_{\hat{\alpha}}\Psi(x)+\Gamma_{\hat{\alpha}}\Psi(x)\,, 
\end{equation}
chosen so that $\nabla_{\hat{\alpha}}\Psi(x)$ transforms cleanly as a spinor:
\begin{equation}
\nabla_{\hat{\alpha}}\Psi(x)\;\to\;S(x)\,\nabla_{\hat{\alpha}}\Psi(x).
\end{equation}
This is exactly the origin of the $\Gamma$ term in  Eq.~(\ref{eq:spincovdif}).
The Fock--Ivanenko coefficients matrices, $\Gamma_{\hat{\alpha}}(x)$, are the orthonormal 
spinor connection. Conceptually, they play exactly the role you already
understand from ordinary moving frames: they supply the ``basis-variation part''
of the derivative. The only novelty is that the basis is now a basis in spinor
space, so the correction must be a $4\times 4$ matrix acting on $\Psi(x)$.

Finally, it is important to see why the orthonormal spinor connection is not simply the
Ricci Rotation coefficients $\omega_{\hat{\alpha}}{}^{\hat{\beta}}{}_{\hat{\gamma}}$. The object $\omega_{\hat{\alpha}}{}^{\hat{\beta}}{}_{\hat{\gamma}}$ acts
on Lorentz-vector indices; it tells you how orthonormal vectors rotate and
boost. A spinor has no such index, so $\omega$ cannot act on it directly.
Instead, the same underlying local Lorentz rotation/boost must be represented
in the space of spinors. Thus $\Gamma_{\hat{\alpha}}$ must be built from the same
geometric information contained in $\omega$, but expressed as a matrix in the
spinor representation. In this way, vectors and spinors are being
parallel-transported by the same local Lorentz geometry; they simply ``feel''
it through different representations.

The orthonormal connection used for spinors, (i.e. the Fock-Ivanenko connection), allowing for spacetime dependent gamma-matrices now, has components which can be calculated in the following way:
\begin{equation}
\boxed{ \Gamma_{\hat{\delta}}(x)=\frac{1}{4}\gamma_{\hat{\alpha}}(x)\left(\partial_{\hat{\delta}}\gamma^{\hat{\alpha}}(x) +{\omega_{\hat{\delta}}^{\;\;\hat{\alpha}}}_{\hat{\beta}}(x) \gamma^{\hat{\beta}}(x)\right)+\; a_{\hat{\delta}}(x)\,\mathbbm{1}}\,, \label{eq:fockivan}
\end{equation}
where $\gamma_{\hat{\alpha}}(x)=\eta_{\hat{\alpha}\hat{\nu}}\gamma^{\hat{\nu}}(x)$ and $a_{\hat{\delta}}(x)=i\,\partial_{\hat{\delta}}\lambda(x)$ with $\lambda(x)$ a real scalar function which yields a pure-gauge
term proportional to the identity (removable by $\Psi(x)\mapsto e^{-i\lambda(x)}\Psi(x)$). Usually the components $a_{\hat{\delta}}(x)$ are taken as zero which is a convention that we shall employ in the rest of this manuscript.

With the choice in Eq.~(\ref{eq:psibar}) the Fock-Ivanenko connection is required to satisfy the ``Hermiticity'' property that
\begin{equation}
\Gamma_{\hat{\alpha}}^{\dagger}(x)
=
-\gamma^{\hat{0}}(x)\,\Gamma_{\hat{\alpha}}(x)\,\gamma^{\hat{0}}(x)
+
\big(\partial_{\hat{\alpha}}\gamma^{\hat{0}}(x)\big)\,\gamma^{\hat{0}}(x)\, .
\label{eq:FIherm2}
\end{equation}
Here we again draw the attention of the reader to the comments after equation Eq.~(\ref{eq:orthgamherm}). With this restriction, the Fock-Ivanenko connection defined in Eq.~(\ref{eq:fockivan})  ensures that the gamma matrices are ``covariantly constant'' with respect to the following covariant derivative
\begin{equation}
\nabla_{\hat{\delta}}\gamma^{\hat{\alpha}}(x)\;:=\;
\partial_{\hat{\delta}}\gamma^{\hat{\alpha}}(x) + \omega_{\hat{\delta}}{}^{\hat{\alpha}}{}_{\hat{\beta}}(x)\, \gamma^{\hat{\beta}}(x) +
\big[\Gamma_{\hat{\delta}}(x),\,\gamma^{\hat{\alpha}}(x)\big]\equiv 0 \,, \label{eq:covdifgamma}
\end{equation}\\[-0.1cm]
with the usual notation for the commutator $[A,\,B]=AB-BA$ (matrix multiplication in this case).

Notice that the matrix structure of Eq.~(\ref{eq:fockivan}) allows it to directly couple to spinors, as it should, via $\Gamma_{\hat{\delta}\,AB}\Psi_{B}(x)$ (sum over $B$, with capital indices denoting spinor component indices), yielding an index structure compatible with the partial derivative term in Eq.~(\ref{eq:arbdireqn}). That is, the Dirac equation in arbitrary coordinate systems, Eq.~(\ref{eq:arbdireqn}), when written in fully component notation, is \\
\begin{equation}
\Bigr[i\gamma^{\hat{\alpha}}_{AB}\left(\delta_{BC}\partial_{\hat{\alpha}} + \Gamma_{\hat{\alpha}\,BC}(x)\right) -m\,\delta_{AC}\Bigr]\Psi_{C}(x)  = 0 \quad\text{(summation over all repeated indices)}\, \label{eq:Deqncomponents}
\end{equation}\\[-0.1cm]
($\delta_{BC}$ is the identity matrix written in component form as a Kronecker-delta).

We now have all the ingredients in order to write down the Dirac equation of
Eq.~(\ref{eq:arbdireqn}) or Eq.~(\ref{eq:Deqncomponents}) in any coordinate system. Further, with our definitions
and conventions, the Dirac action may be written in the manifestly real
(symmetrized) form
\begin{equation}
S[\Psi,\overline{\Psi}]
=
\int d^{4}x\,e\,
\left[
\frac{i}{2}\Big(
\overline{\Psi}\,\gamma^{\hat\alpha}\,\nabla_{\hat\alpha}\Psi
-
(\nabla_{\hat\alpha}\overline{\Psi})\,\gamma^{\hat\alpha}\Psi
\Big)
- m\,\overline{\Psi}\Psi
\right], \label{eq:diract}
\end{equation}
where $e:=\det(e_{\mu}{}^{\hat\alpha})$.
This action is invariant under local Lorentz transformations of the orthonormal
frame. Writing the Lorentz transformation at each spacetime point as
${\Lambda^{\hat\alpha}}_{\hat\beta}(x)\in \text{SO}^{+}(1,3)$, and its spin lift
as $S(\Lambda(x))\in \mathrm{Spin}^{+}(1,3)$, the standard transformation laws
are
\begin{equation}
\Psi'(x)=S(\Lambda(x))\,\Psi(x),
\qquad
\overline{\Psi}'(x)=\overline{\Psi}(x)\,S(\Lambda(x))^{-1}, 
\end{equation}
together with the defining compatibility relation
\begin{equation}
S(\Lambda(x))^{-1}\,\gamma^{\hat\alpha}(x)\,S(\Lambda(x))
=
{\Lambda^{\hat\alpha}}_{\hat\beta}(x)\,\gamma^{\hat\beta}(x).
\end{equation}
The Fock-Ivanenko connection transforms as a gauge field,
\begin{equation}
\Gamma'_{\hat\alpha}(x)
=
S(\Lambda(x))\,\Gamma_{\hat\alpha}(x)\,S(\Lambda(x))^{-1}
-
\big(\partial_{\hat\alpha}S(\Lambda(x))\big)\,S(\Lambda(x))^{-1}.
\end{equation}
Equivalently, these imply the covariant transformation property of the
derivative,
\begin{equation}
\nabla'_{\hat\alpha}\Psi' = S(\Lambda)\,\nabla_{\hat\alpha}\Psi,
\qquad
\nabla'_{\hat\alpha}\overline{\Psi}' = (\nabla_{\hat\alpha}\overline{\Psi})\,S(\Lambda)^{-1},
\end{equation}
so that the integrand in $S[\Psi,\overline{\Psi}]$ is a local Lorentz scalar and the
action is invariant.

We should also mention the Dirac current, which even in this generalized framework has the same structure as its well known form:
\begin{equation}
 J^{\hat{\alpha}}(x) = \overline{\Psi}(x)\gamma^{\hat{\alpha}}(x)\Psi(x)\, \label{eq:dircurrent}
\end{equation}
(note that this current's components \emph{do} generally differ from their holonomic counterparts $J^{\mu}=\overline{\Psi}(x)\gamma^{\mu}(x)\Psi(x)$\,).
Such a current is locally conserved in the standard sense
\begin{equation}
 \nabla_{\hat{\alpha}}J^{\hat{\alpha}}(x) = \left(\nabla_{\hat{\alpha}}\overline{\Psi}(x)\right)\gamma^{\hat{\alpha}}(x)\Psi(x) + \overline{\Psi}(x)\left(\nabla_{\hat{\alpha}}\gamma^{\hat{\alpha}}(x)\right)\Psi(x) + \overline{\Psi}(x)\gamma^{\hat{\alpha}}(x)\Bigl(\nabla_{\hat{\alpha}}\Psi(x)\Bigr)=0. \label{eq:gencurcons}
\end{equation}
Note that the derivatives on the spinors are furnished by Eq.~(\ref{eq:spincovdif}) and Eq.~(\ref{eq:psibarcovdif}), whereas the covariant derivative of the gamma matrix is given by
Eq.~(\ref{eq:covdifgamma}) and therefore vanishes so that the gamma matrix is ``covariantly constant'' yielding
\begin{equation}
 \nabla_{\hat{\alpha}}J^{\hat{\alpha}}(x)=\left(\nabla_{\hat{\alpha}}\overline{\Psi}(x)\right)\gamma^{\hat{\alpha}}(x)\Psi(x) + \overline{\Psi}(x)\gamma^{\hat{\alpha}}(x)\Bigl(\nabla_{\hat{\alpha}}\Psi(x)\Bigr)=0 \label{eq:curcons}
\end{equation}
for the conservation law.

\subsection{The inclusion of gravity}
What we have discussed above already has all the ingredients required to include torsion-free gravity and no additional elements are required. The only new concept we need to discuss here is that the presence of gravity is heralded by the non-vanishing of at least some components of the \emph{Riemann curvature tensor}
\begin{equation}
R_{\hat{\gamma}\hat{\delta}}{}^{\hat{\alpha}}{}_{\hat{\beta}}
=\partial_{\hat{\gamma}}\omega_{\hat{\delta}}{}^{\hat{\alpha}}{}_{\hat{\beta}}
-\partial_{\hat{\delta}}\omega_{\hat{\gamma}}{}^{\hat{\alpha}}{}_{\hat{\beta}}
+\omega_{\hat{\gamma}}{}^{\hat{\alpha}}{}_{\hat{\epsilon}}\,
 \omega_{\hat{\delta}}{}^{\hat{\epsilon}}{}_{\hat{\beta}}
-\omega_{\hat{\delta}}{}^{\hat{\alpha}}{}_{\hat{\epsilon}}\,
 \omega_{\hat{\gamma}}{}^{\hat{\epsilon}}{}_{\hat{\beta}}
-\Big(\omega_{\hat{\gamma}}{}^{\hat{\epsilon}}{}_{\hat{\delta}}
-\omega_{\hat{\delta}}{}^{\hat{\epsilon}}{}_{\hat{\gamma}}\Big)\,
 \omega_{\hat{\epsilon}}{}^{\hat{\alpha}}{}_{\hat{\beta}}
\,.\label{eq:riem}
\end{equation}

In the absence of gravity the coordinate component metric, $g_{\mu\nu}(x)$, which is used in the construction of the tetrad as discussed earlier, must yield Ricci rotation coefficients which give zero for all components of Eq.~(\ref{eq:riem}). On the other hand, if there is gravity, the tetrad, via the Ricci rotation coefficients, must render at least some of the components of Eq.~(\ref{eq:riem}) non-zero.   Aside from this, there is no impediment to minimally coupling gravity to the Dirac equation, and all equations of  Sec.~\ref{sec:covdif} as well as Eq.~(\ref{eq:arbdireqn}) hold. 

One other quantity that is important when considering gravitation is the \emph{stress-energy tensor}. For the Dirac field the most popular form of the symmetric stress-energy tensor, in an orthonormal frame, is given by
\begin{equation}
T_{\hat{\alpha}\hat{\beta}}(x)
=
\frac{i}{2}\Big[
\overline{\Psi}(x)\,\gamma{}_{(\hat \alpha}(x)\,
\nabla_{\hat{\beta}}{}_{)} \Psi(x)
-\big(\nabla_{(}{}_{\hat{\alpha}}\overline{\Psi}(x)\big)\,
\gamma_{\hat{\beta}}{}_{)}(x)\,\Psi(x)
\Big]\,, \label{eq:dirset}
\end{equation}
where the parentheses around the indices denote symmetrization: $X_{(\hat{\alpha}\hat{\beta})}:=\left(X_{\hat{\alpha}\hat{\beta}}+X_{\hat{\beta}\hat{\alpha}}\right)/2$\,. (As with the current, the orthonormal stress-energy tensor's components generally do differ from their holonomic counterparts. The same is true for the Riemann curvature tensor.)

\section{Examples and special cases}\label{sec:examples}
\subsection{Spherical coordinates}\label{subsec:spherical} 
\renewcommand{\aa}{a(t,r)}
\newcommand{\bb}{b(t,r)}
As this is the first example, we will present this particular case with a higher level of detail than the subsequent examples.
Some interesting works in spacetimes exhibiting spherical symmetry
can be found in e.g.\
\cite{villalba94}, \cite{cotaescu97unp}, \cite{cotaescu07mpla},
\cite{finster09}, and \cite{cotaescu16epjc}.
A particularly nice simple example in flat spacetime and spherical coordinates
can be found in Greiner's book \cite{greiner00rqmwe}.

In spherical coordinates the component metric $g_{\mu\nu}(x)$ is furnished by the following line-element:
\begin{equation}
 ds^{2}=g_{\mu\nu}(x)\,\rmd x^{\mu}\rmd x^{\nu} = \aa\,\rmd t^{2} -\bb\,\rmd r^{2} -r^{2}\rmd\theta^{2}-r^{2}\sin^{2}\theta\,\rmd\phi^{2}\,, \label{eq:sphgenmet} 
\end{equation}
so that
\begin{equation}
 [g_{\mu\nu}(x)]=\left[\begin{array}{cccc}
                     \aa & 0 & 0 & 0 \\
                     0 & -\bb & 0 & 0 \\
                     0 & 0 & -r^2 & 0 \\
                     0 & 0 & 0 & -r^{2}\sin^{2}\theta
                     \end{array}\right]\,, \label{eq:sphmet}
\end{equation}
with $\aa$ and $\bb$ positive. If one wanted to enforce flat spacetime (no gravity) $\aa$ and $\bb$ could be set equal to one in order to make Eq.~(\ref{eq:riem}) vanish.
Since spherical coordinates are most useful in problems exhibiting spherical symmetry we have restricted the metric functions $a$ and $b$ to depend only on the coordinates $t$ and $r$.

 For this first example we will choose a tetrad where the four orthonormal tetrad vectors point in the direction of the coordinates. Let us begin with constructing the tetrad vector pointing in the $\phi$ direction, $e_{\mu}^{\;\,\hat{\phi}}$. First we consider a $\phi$=const. level surface defined by the equation
\begin{equation}
 f(\phi)=\phi-\phi_{0}=0\,, \label{eq:philevel}
\end{equation}
with $\phi_{0}$ a constant. Taking the coordinate gradient of this function yields a vector pointing orthogonal to the level surface, $n_{\mu}$:
\begin{equation}
 n_{\mu}=\partial_{\mu}f(\phi)= \delta_{\mu}^{\;\,\phi}\,. \label{eq:nphi}
\end{equation}
Now, although this vector locally points in the $\phi$ direction, it is not normalized to unity. We therefore need to divide it by its magnitude in order to normalize it. This then yields $e_{\mu}^{\;\,\hat\phi}(x)$ as
\begin{equation}
 e_{\mu}^{\;\,\hat\phi}(x)=\frac{n_{\mu}}{\sqrt{|n_{\sigma}g^{\sigma\rho}n_{\rho}|}}=\frac{\delta_{\mu}^{\;\,\phi}}{\sqrt{|\delta_{\sigma}^{\;\,\phi}g^{\sigma\rho}\delta_{\rho}^{\;\,\phi}|}}=\frac{\delta_{\mu}^{\;\,\phi}}{\sqrt{|g^{\phi\phi}|}}=\delta_{\mu}^{\;\,\phi}\, r\sin\theta\,. 
\end{equation}
The normalization is performed with respect to the coordinate metric since the index $\mu$ is a coordinate index, not an orthonormal one.

We can follow a similar procedure for each of the other coordinate directions yielding
\begin{equation}
 e_{\mu}^{\;\,\hat{t}}(x)=\delta_{\mu}^{\;\,t}\,\sqrt{\aa}\,, \quad e_{\mu}^{\;\,\hat{r}}(x)=\delta_{\mu}^{\;\,r}\,\sqrt{\bb}\,, \quad e_{\mu}^{\;\,\hat{\theta}}(x) =\delta_{\mu}^{\;\,\theta}\,r\,.
\end{equation}
In other words, the matrix representing the full tetrad, $e_{\mu}^{\;\,\hat{\alpha}}$, is furnished by
\begin{equation}
\left[e_{\mu}^{\;\,\hat{\alpha}}(x)\right]
=
\begin{array}{c c}
&
\begin{array}{@{}cccc@{}}
\Scale[0.67]{\hat{\alpha}=\hat{0}\;\,(\hat{t})} &
\Scale[0.67]{\hat{\alpha}=\hat{1}\;\,(\hat{r})} &
\Scale[0.67]{\hat{\alpha}=\hat{2}\;\,(\hat{\theta})} &
\Scale[0.67]{\hat{\alpha}=\hat{3}\;\,(\hat{\phi})}
\end{array}
\\
\begin{array}{@{}c@{}}
\Scale[0.70]{\mu=0\;\,(t)} \\
\Scale[0.70]{\mu=1\;\,(r)} \\
\Scale[0.70]{\mu=2\;\,(\theta)} \\
\Scale[0.70]{\mu=3\;\,(\phi)}
\end{array}
&
\left[
\begin{array}{cccc}
\sqrt{\aa} & 0 & 0 & 0 \\
0 & \sqrt{\bb} & 0 & 0 \\
0 & 0 & r & 0 \\
0 & 0 & 0 & r\sin\theta
\end{array}
\right]
\end{array}
\,.
\label{eq:sphorthotet}
\end{equation}

Recalling that orthonormal indices are raised and lowered with the orthonormal metric of Eq.~(\ref{eq:orthmet}), and coordinate indices are raised and lowered with the coordinate metric of Eq.~(\ref{eq:sphmet}) we also have
\begin{subequations}
\begin{align}
\left[e^{\nu\hat{\alpha}}(x)\right]& = \left[g^{\nu\mu}(x)e_{\mu}^{\;\,\hat{\alpha}}(x)\right] =  \left[
 \begin{array}{c c c c}
  \frac{1}{\sqrt{\aa}} & 0 & 0 &0 \\
  0 & -\frac{1}{\sqrt{\bb}} & 0 & 0 \\
  0 & 0 & -\frac{1}{r} & 0 \\
  0 & 0 & 0 & -\frac{1}{r\sin\theta}
 \end{array}\right], \label{eq:sphtetsi} \\[0.2cm] 
 \left[e^{\nu}_{\;\hat{\beta}}(x)\right] & = \left[g^{\nu\mu}(x) e_{\mu}^{\;\,\hat{\alpha}}(x) \eta_{\hat{\alpha}\hat{\beta}} \right] = \left[
 \begin{array}{c c c c}
   \frac{1}{\sqrt{\aa}} & 0 & 0 &0 \\
  0 & \frac{1}{\sqrt{\bb}} & 0 & 0 \\
  0 & 0 & \frac{1}{r} & 0 \\
  0 & 0 & 0 & \frac{1}{r\sin\theta}
 \end{array}\right], \label{eq:sphtetsii} \\[0.2cm]
\left[e_{\mu\hat{\alpha}}(x)\right] & =  \left[e_{\mu}^{\;\,\hat{\alpha}}(x) \eta_{\hat{\alpha}\hat{\beta}} \right] = \left[
 \begin{array}{c c c c}
 \sqrt{\aa} & 0 & 0 &0 \\
  0 & -\sqrt{\bb} & 0 & 0 \\
  0 & 0 & -r & 0 \\
  0 & 0 & 0 & -r\sin\theta
 \end{array}\right]. \label{eq:sphtetsiii}
\end{align}
\end{subequations}
From these tetrads we can construct the non-zero Ricci rotation coefficients from Eq.~(\ref{eq:rrcformula}) as:
\begin{equation}
 \omega_{\hat{0}}^{\;\,\hat{1}\hat{0}}=\frac{\partial_{r}\aa}{2\aa\sqrt{\bb}}\,, \quad  \omega_{\hat{1}}^{\;\,\hat{1}\hat{0}}=\frac{\partial_{t}\bb}{2\sqrt{\aa}\bb}\,, \quad \omega_{\hat{2}}^{\;\,\hat{1}\hat{2}}=\frac{1}{r\sqrt{\bb}}=\omega_{\hat{3}}^{\;\,\hat{1}\hat{3}}\,, \quad \omega_{\hat{3}}^{\;\,\hat{2}\hat{3}}=\frac{\cot\theta}{r}\,. \label{eq:sphrrcs}
\end{equation}
as well as those related to the above via antisymmetry on the last two indices.
Here we will take the orthonormal gamma matrices to be the common Pauli-Dirac ones in Eq.~(\ref{eq:paulidiracgams1}) and Eq.~(\ref{eq:paulidiracgams2}). With these ingredients we construct the Fock-Ivanenko connection of Eq.~(\ref{eq:fockivan}) required for the Dirac equation:
\allowdisplaybreaks{\begin{align}
\Gamma_{\hat{0}}& =\left[\begin{array}{cccc}
                        0 & 0 & 0 & \frac{\partial_{r}\aa}{4\aa\sqrt{\bb}} \\
                        0 & 0 &  \frac{\partial_{r}\aa}{4\aa\sqrt{\bb}} & 0 \\
                        0 &  \frac{\partial_{r}\aa}{4\aa\sqrt{\bb}} & 0 & 0 \\
                        \frac{\partial_{r}\aa}{4\aa\sqrt{\bb}} & 0 & 0 &0
                       \end{array}\right]\,, \nonumber \\
\Gamma_{\hat{1}}& = \left[\begin{array}{cccc}
                        0 & 0 & 0 & \frac{\partial_{t}\bb}{4\sqrt{\aa}\bb} \\
                        0 & 0 &  \frac{\partial_{t}\bb}{4\sqrt{\aa}\bb} & 0 \\
                        0 &  \frac{\partial_{t}\bb}{4\sqrt{\aa}\bb} & 0 & 0 \\
                        \frac{\partial_{t}\bb}{4\sqrt{\aa}\bb} & 0 & 0 &0
                       \end{array}\right]\,,\nonumber \\
\Gamma_{\hat{2}}& =\left[\begin{array}{cccc}
                        -\frac{i}{2r\sqrt{\bb}} & 0 & 0 & 0 \\
                        0 & \frac{i}{2r\sqrt{\bb}} &  0 & 0 \\
                        0 &  0 & -\frac{i}{2r\sqrt{\bb}} & 0 \\
                        0 & 0 & 0 &\frac{i}{2r\sqrt{\bb}}
                       \end{array}\right]\, , \nonumber  \\
\Gamma_{\hat{3}} & = \left[\begin{array}{cccc}
                        0 & \frac{\frac{1}{\sqrt{\bb}} - i \cot\theta}{2r} & 0 & 0 \\
                        -\frac{\frac{1}{\sqrt{\bb}} - i \cot\theta}{2r} & 0 &  0 & 0 \\
                        0 &  0 & 0 & \frac{\frac{1}{\sqrt{\bb}} - i \cot\theta}{2r} \\
                        0 & 0 & -\frac{\frac{1}{\sqrt{\bb}} - i \cot\theta}{2r} &0
                       \end{array}\right]\,.\label{eq:sphrFIs}
\end{align}}

Therefore, the Dirac equation of Eq.~(\ref{eq:arbdireqn}) or Eq.~(\ref{eq:Deqncomponents}) boils down to the following set:
\begin{subequations}
\begin{align}
0=&\frac{1}{4 r\,\aa\,\bb}\Bigg(
\,4 i r \sqrt{\aa}\,\bb\,\partial_{t}\Psi_{1}(x)
+4 i\,\aa\,r\sqrt{\bb}\,\partial_{r}\Psi_{4}(x)
+4\,\aa\bb\,\partial_{\theta}\Psi_{4}(x) \nonumber\\
&\quad
+4 i\,\aa\bb\,\csc\theta\,\partial_{\phi}\Psi_{3}(x)
-4 m r\,\aa\bb\,\Psi_{1}(x)
+2\,\aa\bb\,\cot\theta\,\Psi_{4}(x) \nonumber\\
&\quad
+4 i\,\aa\sqrt{\bb}\,\Psi_{4}(x)
+i r \sqrt{\bb}\,\Psi_{4}(x)\,\partial_{r}\aa
+i r \sqrt{\aa}\,\Psi_{1}(x)\,\partial_{t}\bb
\Bigg),  \label{eq:sphdireqn1} \\[0.2cm]
0=&\frac{1}{4 r\,\aa\,\bb}\Bigg(
\,4 i r \sqrt{\aa}\,\bb\,\partial_{t}\Psi_{2}(x)
+4 i\,\aa\,r\sqrt{\bb}\,\partial_{r}\Psi_{3}(x)
-4\,\aa\bb\,\partial_{\theta}\Psi_{3}(x) \nonumber\\
&\quad
-4 i\,\aa\bb\,\csc\theta\,\partial_{\phi}\Psi_{4}(x)
-4 m r\,\aa\bb\,\Psi_{2}(x)
-2\,\aa\bb\,\cot\theta\,\Psi_{3}(x) \nonumber\\
&\quad
+4 i\,\aa\sqrt{\bb}\,\Psi_{3}(x)
+i r \sqrt{\bb}\,\Psi_{3}(x)\,\partial_{r}\aa
+i r \sqrt{\aa}\,\Psi_{2}(x)\,\partial_{t}\bb
\Bigg), \label{eq:sphdireqn2} \\[0.2cm]
0=&\frac{1}{4 r\,\aa\,\bb}\Bigg(
\,4 i r \sqrt{\aa}\,\bb\,\partial_{t}\Psi_{3}(x)
+4 i\,\aa\,r\sqrt{\bb}\,\partial_{r}\Psi_{2}(x)
+4\,\aa\bb\,\partial_{\theta}\Psi_{2}(x) \nonumber\\
&\quad
+4 i\,\aa\bb\,\csc\theta\,\partial_{\phi}\Psi_{1}(x)
+4 m r\,\aa\bb\,\Psi_{3}(x)
+2\,\aa\bb\,\cot\theta\,\Psi_{2}(x) \nonumber\\
&\quad
+4 i\,\aa\sqrt{\bb}\,\Psi_{2}(x)
+i r \sqrt{\bb}\,\Psi_{2}(x)\,\partial_{r}\aa
+i r \sqrt{\aa}\,\Psi_{3}(x)\,\partial_{t}\bb
\Bigg) , \label{eq:sphdireqn3} \\[0.2cm]
0=&\frac{1}{4 r\,\aa\,\bb}\Bigg(
-\,4 i r \sqrt{\aa}\,\bb\,\partial_{t}\Psi_{4}(x)
+4 i\,\aa\,r\sqrt{\bb}\,\partial_{r}\Psi_{1}(x)
+4\,\aa\bb\,\partial_{\theta}\Psi_{1}(x) \nonumber\\
&\quad
+4 i\,\aa\bb\,\csc\theta\,\partial_{\phi}\Psi_{2}(x)
-4 m r\,\aa\bb\,\Psi_{4}(x)
+2\,\aa\bb\,\cot\theta\,\Psi_{1}(x) \nonumber\\
&\quad
-4 i\,\aa\sqrt{\bb}\,\Psi_{1}(x)
- i r \sqrt{\bb}\,\Psi_{1}(x)\,\partial_{r}\aa
- i r \sqrt{\aa}\,\Psi_{4}(x)\,\partial_{t}\bb
\Bigg), \label{eq:sphdireqn4}
\end{align}
\end{subequations}
where the $\Psi_{A}(x)$ denote the components of the bispinor:
\begin{equation}
 \Psi=\left[\begin{array}{c}
             \Psi_{1}(x)\\
             \Psi_{2}(x)\\
             \Psi_{3}(x)\\
             \Psi_{4}(x)\\
            \end{array} \right]\,. \nonumber
\end{equation}

Of particular importance in the realm of gravity with spherical symmetry is when the spacetime is a priori set to the \emph{Schwarzschild  spacetime}. For this (time independent) scenario we have that in Eq.~(\ref{eq:sphmet})
\begin{equation}
 a(r)=1-\frac{2M}{r} \qquad \text{and} \qquad b(r)=\frac{1}{1-\frac{2M}{r}}\,, \label{eq:schwmet}
\end{equation}
where $M$ represents the mass of the gravitating body and here we assume $r>2M$. For this specific scenario, using Eqs.~(\ref{eq:sphtetsi}-c) and Eq.~(\ref{eq:paulidiracgams2}), the Dirac equations of Eq.~(\ref{eq:sphdireqn1} - \ref{eq:sphdireqn4}) reduce  to:
\begin{subequations}
\begin{align}
0=&\frac{2 i r}{\sqrt{1-\frac{2M}{r}}}\,\partial_{t}\Psi_{1}(x)
+2 i r \sqrt{1-\frac{2M}{r}}\,\partial_{r}\Psi_{4}(x)
+2\,\partial_{\theta}\Psi_{4}(x) \nonumber\\
&\quad
+2 i\,\csc\theta\,\partial_{\phi}\Psi_{3}(x)
-2 m r\,\Psi_{1}(x)
+\cot\theta\,\Psi_{4}(x)
+\frac{i(2r-3M)}{r\sqrt{1-\frac{2M}{r}}}\,\Psi_{4}(x)\,,
\label{eq:schdeq1} \\[0.1cm]
0=&\frac{2 i r}{\sqrt{1-\frac{2M}{r}}}\,\partial_{t}\Psi_{2}(x)
+2 i r \sqrt{1-\frac{2M}{r}}\,\partial_{r}\Psi_{3}(x)
-2\,\partial_{\theta}\Psi_{3}(x) \nonumber\\
&\quad
-2 i\,\csc\theta\,\partial_{\phi}\Psi_{4}(x)
-2 m r\,\Psi_{2}(x)
-\cot\theta\,\Psi_{3}(x)
+\frac{i(2r-3M)}{r\sqrt{1-\frac{2M}{r}}}\,\Psi_{3}(x)\,,
\label{eq:schdeq2}\\[0.1cm]
0=&\frac{2 i r}{\sqrt{1-\frac{2M}{r}}}\,\partial_{t}\Psi_{3}(x)
+2 i r \sqrt{1-\frac{2M}{r}}\,\partial_{r}\Psi_{2}(x)
+2\,\partial_{\theta}\Psi_{2}(x) \nonumber\\
&\quad
+2 i\,\csc\theta\,\partial_{\phi}\Psi_{1}(x)
+2 m r\,\Psi_{3}(x)
+\cot\theta\,\Psi_{2}(x)
+\frac{i(2r-3M)}{r\sqrt{1-\frac{2M}{r}}}\,\Psi_{2}(x)\,,
\label{eq:schdeq3} \\[0.1cm]
0=&-\frac{2 i r}{\sqrt{1-\frac{2M}{r}}}\,\partial_{t}\Psi_{4}(x)
+2 i r \sqrt{1-\frac{2M}{r}}\,\partial_{r}\Psi_{1}(x)
+2\,\partial_{\theta}\Psi_{1}(x) \nonumber\\
&\quad
+2 i\,\csc\theta\,\partial_{\phi}\Psi_{2}(x)
-2 m r\,\Psi_{4}(x)
+\cot\theta\,\Psi_{1}(x)
-\frac{i(2r-3M)}{r\sqrt{1-\frac{2M}{r}}}\,\Psi_{1}(x)\,. \label{eq:schdeq4}
\end{align}
\end{subequations}
Some analyses and approximate solutions of the Dirac in Schwarzschild spacetime may be found in \cite{mukhopadhyay99cqg}, \cite{lyu07ps}, \cite{cotaescu07xwmpla}, \cite{alsing22arx}, \cite{calvomunoz25}.

Before going on to the next example, it is worth mentioning a few other special cases in the realm of spherically symmetric spacetimes. If instead of the tetrad in Eq.~(\ref{eq:sphorthotet}) one takes the following tetrad:
\begin{equation}
\left[e_{\mu}^{\;\hat{\alpha}}(x)\right]=\left[
 \begin{array}{c c c c}
  \sqrt{\aa} & 0 & 0 &0 \\
  0 & \sqrt{\bb}\sin\theta\,\cos\phi & \sqrt{\bb}\sin\theta\,\sin\phi & \sqrt{\bb}\cos\theta \\
  0 & r\cos\theta\,\cos\phi & r\cos\theta\,\sin\phi & -r\sin\theta \\
  0 & -r\sin\theta\,\sin\phi & r\sin\theta\,\cos\phi & 0
 \end{array}\right]\,, \label{eq:rottet}
\end{equation}
which still obeys Eq.~(\ref{eq:tetradorthonormal}) and Eq.~(\ref{eq:metricfromtetrad}), one gets \emph{zero} for all components of the Ricci rotation coefficients $\omega_{\hat{\delta}}^{\;\,\hat{\alpha}\hat{\beta}}$ in the flat spacetime limit (i.e. no gravity, $\aa=1$, $\bb=1$). Further, if along with this tetrad one picks  a constant set of orthonormal gamma matrices, $\gamma^{\hat{\alpha}}$, the Fock-Ivanenko connection also vanishes in the absence of gravity. However, the price to pay is that the orthonormal derivative $\partial_{\hat{\alpha}}=e^{\mu}_{\;\,\hat{\alpha}}(x)\,\partial_{\mu}$ in the Dirac equation is more complicated.

Another interesting case is the following: Consider a ``spherical projection of the Pauli matrices'' furnished by:
\begin{align}
\sigma^{r}&:=\bm{\sigma}\cdot \orthvec{\hat{r}} = 
\left(\begin{array}{cc}
       \cos\theta & e^{-i\phi}\sin\theta \\
       e^{i\phi}\sin\theta & -\cos\theta
      \end{array}\right), \;
\sigma^{\theta}:=\bm{\sigma}\cdot \orthvec{\hat{\theta}} = 
\left(\begin{array}{cc}
       -\sin\theta & e^{-i\phi}\cos\theta \\
       e^{i\phi}\cos\theta & \sin\theta
      \end{array}\right), \;  \nonumber \\
\sigma^{\phi}&:=\bm{\sigma}\cdot \orthvec{\hat{\phi}} =
\left(\begin{array}{cc}
       0 & -ie^{-i\phi} \\
       ie^{i\phi} & 0
      \end{array}\right), \label{eq:spsigmas}
\end{align}
with $\bm{\sigma}$ the {Pauli vector} $\bm{\sigma}:=[\sigma^{x},\,\sigma^{y},\,\sigma^{z}]$ and $\orthvec{\hat{j}}$ unit vectors pointing in the $j$ direction:
\begin{align}
 \orthvec{\hat{r}}&=\sin\theta\cos\phi\,\hat{\text{\bf x}} + \sin\theta\sin\phi\,\hat{\text{\bf y}}+\cos\theta \,\hat{\text{\bf z}}\,, \nonumber \\
 \orthvec{\hat{\theta}}&=\cos\theta\cos\phi\,\hat{\text{\bf x}} + \cos\theta\sin\phi\,\hat{\text{\bf y}}-\sin\theta \,\hat{\text{\bf z}}\,, \nonumber \\
 \orthvec{\hat{\phi}}&=-\sin\phi\,\hat{\text{\bf x}} + \cos\phi\,\hat{\text{\bf y}}\,. \nonumber
\end{align}

From Eq.~(\ref{eq:spsigmas}) a set of ``spherical orthonormal gamma matrices'' can be constructed as
\begin{equation}
\gamma^{\hat{0}}=\left[\begin{array}{cc}
                        \mathbbm{1} & \mathbbm{0} \\
                        \mathbbm{0} & -\mathbbm{1}
                       \end{array}\right]\,, \quad
\gamma^{\hat{r}}=\left[\begin{array}{cc}
                        \mathbbm{0} & \sigma^{r} \\
                        -\sigma^{r} & \mathbbm{0}
                       \end{array}\right]\,, \quad
\gamma^{\hat{\theta}}=\left[\begin{array}{cc}
                        \mathbbm{0} & \sigma^{\theta} \\
                        -\sigma^{\theta} & \mathbbm{0}
                       \end{array}\right]\,, \quad
\gamma^{\hat{\phi}}=\left[\begin{array}{cc}
                        \mathbbm{0} & \sigma^{\phi} \\
                        -\sigma^{\phi} & \mathbbm{0}
                       \end{array}\right]\,,
                       \label{eq:sphdiracgams}
\end{equation}
This set of gamma matrices still obeys the Clifford bracket in Eq.~(\ref{eq:clifalg}) but now possesses non-constant entries. It can be checked by brute-force computation that, with the spherical tetrad of Eq.~(\ref{eq:sphorthotet}), and in flat spacetime (no gravity), this choice also yields zero for the Fock-Ivanenko connection, although it does \emph{not} yield zero for the spin-connection or Ricci rotation coefficients. (Note that now the derivative of $\gamma$  in Eq.~(\ref{eq:fockivan}) contributes due to the non-constancy of the gamma matrices. This causes cancellation with the Ricci rotation coefficient term, yielding zero.)

\subsection{Cylindrical coordinates}
\newcommand{\cc}{c(t,r)}
The calculational method here is similar to the spherical coordinates section so the presentation here will be briefer than what we had there. 
Works considering Dirac equation in cylindrical symmetry, especially in the astrophysically interesting scenarios of cosmic string spacetimes, are e.g.\
\cite{mihaila06prd}, \cite{bakke10prd}, \cite{bakke12epjp},
\cite{andrade14epjc}, \cite{guvendi22dark},
\cite{oliveira22grg}, and \cite{boumali26npb}.

In cylindrical coordinates, for the purposes of this example, we take the spacetime metric as
\begin{equation}
 ds^{2}=g_{\mu\nu}\,\rmd x^{\mu}\rmd x^{\nu} = \aa\,\rmd t^{2} - \bb\,\rmd r^{2} - r^{2}\rmd\phi^{2} - \cc\,\rmd z^{2}\,, \label{eq:cylgenmet}
\end{equation}
where the coordinate dependence to $t$ and $r$ only in the metric functions is used because we are assuming that the problem to be solved is cylindrically symmetric. In the scenario where there is no gravity, one would take $\aa=1$, $\bb=1$ and $\cc=1$ which makes all components of the Riemann curvature tensor in Eq.~(\ref{eq:riem}) vanish. 

Following the normal vector procedure of Sec.~\ref{subsec:spherical} but with the metric of Eq.~(\ref{eq:cylgenmet}),  we will take as our tetrad the following:
\begin{equation}
\left[e_{\mu}^{\;\,\hat{\alpha}}(x)\right]=
\left[
\begin{array}{cccc}
\sqrt{\aa} & 0 & 0 & 0 \\
0 & \sqrt{\bb} & 0 & 0 \\
0 & 0 & r & 0 \\
0 & 0 &0 & \sqrt{\cc}
\end{array}\right]\,,\label{eq:cylthotet}
\end{equation}
which, via Eq.~(\ref{eq:rrcformula}), yields the following non-zero components for the Ricci rotation coefficients:
\begin{equation}
\Scale[0.98]{\omega_{\hat{0}}^{\;\,\hat{1}\hat{0}}=\frac{\partial_{r}\aa}{2\aa\sqrt{\bb}},\; \omega_{\hat{1}}^{\;\,\hat{1}\hat{0}}=\frac{\partial_{t}\bb}{2\sqrt{\aa}\bb}, \, \omega_{\hat{2}}^{\;\,\hat{1}\hat{2}}=\frac{1}{r\sqrt{\bb}},\, \omega_{\hat{3}}^{\;\,\hat{3}\hat{0}}=\frac{\partial_{t}\cc}{2\sqrt{\aa}\cc},\, \omega_{\hat{3}}^{\;\,\hat{1}\hat{3}}=\frac{\partial_{r}\cc}{2\sqrt{\bb}\cc}\,.} \label{eq:cylrrcs}
\end{equation}
along with those related to the above by antisymmetry on the last two indices.

For this example we will again choose the Pauli-Dirac representation of Eq.~(\ref{eq:paulidiracgams2}) so that, from Eq.~(\ref{eq:fockivan}) we obtain the following Fock-Ivanenko connection matrices:
\allowdisplaybreaks{\begin{align}
\Gamma_{\hat{0}}&=
\left[
\begin{smallmatrix}
0 & 0 & 0 & \frac{\partial_{r}\aa}{4\,\aa\,\sqrt{\bb}}\\[0.4ex]
0 & 0 & \frac{\partial_{r}\aa}{4\,\aa\,\sqrt{\bb}} & 0\\[0.4ex]
0 & \frac{\partial_{r}\aa}{4\,\aa\,\sqrt{\bb}} & 0 & 0\\[0.4ex]
\frac{\partial_{r}\aa}{4\,\aa\,\sqrt{\bb}} & 0 & 0 & 0
\end{smallmatrix}
\right]\,,\nonumber\\[0.1cm]
\Gamma_{\hat{1}}&=
\left[
\begin{smallmatrix}
0 & 0 & 0 & \frac{\partial_{t}\bb}{4\,\sqrt{\aa}\,\bb}\\[0.4ex]
0 & 0 & \frac{\partial_{t}\bb}{4\,\sqrt{\aa}\,\bb} & 0\\[0.4ex]
0 & \frac{\partial_{t}\bb}{4\,\sqrt{\aa}\,\bb} & 0 & 0\\[0.4ex]
\frac{\partial_{t}\bb}{4\,\sqrt{\aa}\,\bb} & 0 & 0 & 0
\end{smallmatrix}
\right] \,,\nonumber\\[0.1cm]
\Gamma_{\hat{2}}&=
\left[
\begin{smallmatrix}
-\frac{i}{2\,r\,\sqrt{\bb}} & 0 & 0 & 0\\[0.4ex]
0 & \frac{i}{2\,r\,\sqrt{\bb}} & 0 & 0\\[0.4ex]
0 & 0 & -\frac{i}{2\,r\,\sqrt{\bb}} & 0\\[0.4ex]
0 & 0 & 0 & \frac{i}{2\,r\,\sqrt{\bb}}
\end{smallmatrix}
\right]\,,\nonumber\\[0.1cm]
\Gamma_{\hat{3}}&=
\left[
\begin{smallmatrix}
0 &
\frac{\partial_{r}\cc}{4\,\sqrt{\bb}\,\cc} &
\frac{\partial_{t}\cc}{4\,\sqrt{\aa}\,\cc} &
0
\\[0.4ex]
-\frac{\partial_{r}\cc}{4\,\sqrt{\bb}\,\cc} &
0 & 0 &
-\frac{\partial_{t}\cc}{4\,\sqrt{\aa}\,\cc}
\\[0.4ex]
\frac{\partial_{t}\cc}{4\,\sqrt{\aa}\,\cc} &
0 & 0 &
\frac{\partial_{r}\cc}{4\,\sqrt{\bb}\,\cc}
\\[0.4ex]
0 &
-\frac{\partial_{t}\cc}{4\,\sqrt{\aa}\,\cc} &
-\frac{\partial_{r}\cc}{4\,\sqrt{\bb}\,\cc} &
0
\end{smallmatrix}
\right]\,. \label{eq:cylFI}
\end{align}}

Finally, from Eq.~(\ref{eq:arbdireqn}) we have the Dirac equation in cylindrical coordinates:
\begin{subequations}
\begin{align}
0=&\frac{1}{4 r\,\aa\,\bb\,\cc}\Bigg(
 4 i r \sqrt{\aa}\,\bb\,\cc\,\partial_{t}\Psi_{1}(x)
 +4 i\,\aa\,r\sqrt{\bb}\,\cc\,\partial_{r}\Psi_{4}(x)
\nonumber\\
&\qquad
 +4\,\aa\bb\,\cc\,\partial_{\phi}\Psi_{4}(x)
 +4 i r\,\aa\bb\,\sqrt{\cc}\,\partial_{z}\Psi_{3}(x)
\nonumber\\
&\qquad
 -4 m r\,\aa\bb\,\cc\,\Psi_{1}(x)
 + i r \sqrt{\bb}\,\cc\,\Psi_{4}(x)\,\partial_{r}\aa
\nonumber\\
&\qquad
 + i r \sqrt{\bb}\,\aa\,\Psi_{4}(x)\,\partial_{r}\cc
 +2 i\,\aa\sqrt{\bb}\,\cc\,\Psi_{4}(x)
\nonumber\\
&\qquad
 + i r \sqrt{\aa}\,\bb\,\Psi_{1}(x)\,\partial_{t}\cc
 + i r \sqrt{\aa}\,\cc\,\Psi_{1}(x)\,\partial_{t}\bb
\Bigg)\,, \label{eq:dircyl1}
\\[0.8ex]
0=&\frac{1}{4 r\,\aa\,\bb\,\cc}\Bigg(
 4 i r \sqrt{\aa}\,\bb\,\cc\,\partial_{t}\Psi_{2}(x)
 +4 i\,\aa\,r\sqrt{\bb}\,\cc\,\partial_{r}\Psi_{3}(x)
\nonumber\\
&\qquad
 -4\,\aa\bb\,\cc\,\partial_{\phi}\Psi_{3}(x)
 -4 i r\,\aa\bb\,\sqrt{\cc}\,\partial_{z}\Psi_{4}(x)
\nonumber\\
&\qquad
 -4 m r\,\aa\bb\,\cc\,\Psi_{2}(x)
 + i r \sqrt{\bb}\,\cc\,\Psi_{3}(x)\,\partial_{r}\aa
\nonumber\\
&\qquad
 + i r \sqrt{\bb}\,\aa\,\Psi_{3}(x)\,\partial_{r}\cc
 +2 i\,\aa\sqrt{\bb}\,\cc\,\Psi_{3}(x)
\nonumber\\
&\qquad
 + i r \sqrt{\aa}\,\bb\,\Psi_{2}(x)\,\partial_{t}\cc
 + i r \sqrt{\aa}\,\cc\,\Psi_{2}(x)\,\partial_{t}\bb
\Bigg)\,, \label{eq:dircyl2} \\
0=&-\frac{1}{4 r\,\aa\,\bb\,\cc}\Bigg(
 4 i r \sqrt{\aa}\,\bb\,\cc\,\partial_{t}\Psi_{3}(x)
 +4 i\,\aa\,r\sqrt{\bb}\,\cc\,\partial_{r}\Psi_{2}(x)
\nonumber\\
&\qquad
 +4\,\aa\bb\,\cc\,\partial_{\phi}\Psi_{2}(x)
 +4 i r\,\aa\bb\,\sqrt{\cc}\,\partial_{z}\Psi_{1}(x)
\nonumber\\
&\qquad
 +4 m r\,\aa\bb\,\cc\,\Psi_{3}(x)
 +2 i\,\aa\sqrt{\bb}\,\cc\,\Psi_{2}(x)
\nonumber\\
&\qquad
 + i r \sqrt{\bb}\,\cc\,\Psi_{2}(x)\,\partial_{r}\aa
 + i r\,\aa\sqrt{\bb}\,\Psi_{2}(x)\,\partial_{r}\cc
\nonumber\\
&\qquad
 + i r \sqrt{\aa}\,\bb\,\Psi_{3}(x)\,\partial_{t}\cc
 + i r \sqrt{\aa}\,\cc\,\Psi_{3}(x)\,\partial_{t}\bb
\Bigg)\,, \label{eq:dircyl3} \\
0=&-\frac{i}{4 r\,\aa\,\bb\,\cc}\Bigg(
 4 r \sqrt{\aa}\,\bb\,\cc\,\partial_{t}\Psi_{4}(x)
 +4 r\,\aa\sqrt{\bb}\,\cc\,\partial_{r}\Psi_{1}(x)
\nonumber\\
&\qquad
 +4 i\,\aa\bb\cc\,\partial_{\phi}\Psi_{1}(x)
 -4 r\,\aa\bb\,\sqrt{\cc}\,\partial_{z}\Psi_{2}(x)
\nonumber\\
&\qquad
 -4 i m r\,\aa\bb\cc\,\Psi_{4}(x)
 +2\,\aa\sqrt{\bb}\,\cc\,\Psi_{1}(x)
\nonumber\\
&\qquad
 + r \sqrt{\bb}\,\cc\,\Psi_{1}(x)\,\partial_{r}\aa
 + r\,\aa\sqrt{\bb}\,\Psi_{1}(x)\,\partial_{r}\cc
\nonumber\\
&\qquad
 + r\sqrt{\aa}\,\bb\,\Psi_{4}(x)\,\partial_{t}\cc
 + r\sqrt{\aa}\,\cc\,\Psi_{4}(x)\,\partial_{t}\bb
\Bigg)\,.  \label{eq:dircyl4}
\end{align}
\end{subequations}

\subsection{Kerr spacetime}
In this example we study a very important solution to the Einstein field
equations known as the \emph{Kerr solution}
\cite{kerr63prl},\cite{visser07kerr}.
In Boyer-Lindquist coordinates
(a set of oblate spheroidal type coordinates which we denote as $(t,r,\theta,\phi)$)
the line element can be written as
\begin{equation}
\begin{aligned}
\rmd s^{2}={}&\left(1-\frac{2Mr}{\Sigma}\right)\rmd t^{2}
+\frac{4aMr\sin^{2}\theta}{\Sigma}\,\rmd t\,\rmd\phi
-\frac{\Sigma}{\Delta}\,\rmd r^{2}
-\Sigma\,\rmd\theta^{2} \\
&\quad
-\sin^{2}\theta\left(r^{2}+a^{2}
+\frac{2a^{2}Mr\sin^{2}\theta}{\Sigma}\right)\rmd\phi^{2}\,. \label{eq:kerrmet}
\end{aligned}
\end{equation}
Here we have used the standard Kerr abbreviations
\begin{equation}
\Sigma:=r^{2}+a^{2}\cos^{2}\theta,\qquad
\Delta:=r^{2}-2Mr+a^{2}\,,\qquad
\rho:=\sqrt{\Sigma}\,,
\end{equation}
where $M$ is the mass parameter and $a=J/M$ is the spin parameter (i.e. it is the angular momentum per unit mass). We assume here that $r>M+\sqrt{M^{2}-a^{2}}$ with $M^{2}>a^{2}$.

A useful (co)tetrad often used in the studies of the Kerr metric
is the following:
\begin{equation}
 [e_{\mu}^{\;\,\hat{\alpha}}]
 =
 \left[
 \begin{array}{cccc}
  \sqrt{\dfrac{\Delta}{\Sigma}} & 0 & 0 &
   -\,\dfrac{a\sin\theta}{\sqrt{\Sigma}} \\[1.0ex]
  0 & \sqrt{\dfrac{\Sigma}{\Delta}} & 0 & 0 \\[1.0ex]
  0 & 0 & \sqrt{\Sigma} & 0 \\[1.0ex]
  -\,a\sin^{2}\theta\,\sqrt{\dfrac{\Delta}{\Sigma}} & 0 & 0 &
   \dfrac{(r^{2}+a^{2})\sin\theta}{\sqrt{\Sigma}}
 \end{array}
 \right]\,. \label{eq:kerrtet}
\end{equation}
This choice of tetrad is sometimes referred to as the Carter tetrad. It is often used because it puts the Dirac equation into a comparatively simple form and is well suited to separating the equation into radial and angular parts \cite{carter68CMP}, \cite{dolan15cqg}.
Utilizing this tetrad in Eq.~(\ref{eq:rrcformula}) yields the following non-zero components for the Ricci rotation coefficients, plus those related via antisymmetry on the last two indices:
{\allowdisplaybreaks
\begin{align}
\omega_{\hat0}{}^{\hat0\hat1}
&=\frac{a^{2}r\sin^{2}\theta
      -M\!\left(r^{2}-a^{2}\cos^{2}\theta\right)}
      {\rho^{3}\sqrt{\Delta}},
&
\omega_{\hat3}{}^{\hat2\hat3}
&=\frac{(r^{2}+a^{2})\cot\theta}{\rho^{3}},
\nonumber\\[0.8ex]
\omega_{\hat0}{}^{\hat0\hat2}
&=\omega_{\hat1}{}^{\hat1\hat2}
=\frac{a^{2}\sin\theta\cos\theta}{\rho^{3}},
&
\omega_{\hat0}{}^{\hat1\hat3}
&=\omega_{\hat1}{}^{\hat0\hat3}
=\omega_{\hat3}{}^{\hat0\hat1}
=\frac{ar\sin\theta}{\rho^{3}},
\label{eq:kerrrrcs}\\[0.8ex]
\omega_{\hat0}{}^{\hat2\hat3}
&=\omega_{\hat3}{}^{\hat0\hat2}
=-\,\omega_{\hat2}{}^{\hat0\hat3}
=\frac{a\sqrt{\Delta}\cos\theta}{\rho^{3}},
&
\omega_{\hat2}{}^{\hat1\hat2}
&=\omega_{\hat3}{}^{\hat1\hat3}
=\frac{r\sqrt{\Delta}}{\rho^{3}}.
\nonumber
\end{align}
}

Next, using Eq.~(\ref{eq:paulidiracgams2}) and Eq.~(\ref{eq:kerrrrcs}) in Eq.~(\ref{eq:fockivan}) one can find, after long computation,
\begin{align}
\Gamma_{\hat0}
&=\frac12\left[
\begin{array}{cccc}
0 & u-i v & 0 & -q+i p \\
-(u+i v) & 0 & -q-i p & 0 \\
0 & -q+i p & 0 & u-i v \\
-q-i p & 0 & -(u+i v) & 0
\end{array}
\right], \quad
\Gamma_{\hat1}
=\frac12\left[
\begin{array}{cccc}
-i p & 0 & -u & 0 \\
0 & i p & 0 & u \\
-u & 0 & -i p & 0 \\
0 & u & 0 & i p
\end{array}
\right], \nonumber \\[1.0ex]
\Gamma_{\hat2}
&=\frac12\left[
\begin{array}{cccc}
-i s & 0 & v & 0 \\
0 & i s & 0 & -v \\
v & 0 & -i s & 0 \\
0 & -v & 0 & i s
\end{array}
\right], \quad 
\Gamma_{\hat3}
=\frac12\left[
\begin{array}{cccc}
0 & s-i w & 0 & -u+i v \\
-(s+i w) & 0 & -u-i v & 0 \\
0 & -u+i v & 0 & s-i w \\
-u-i v & 0 & -(s+i w) & 0
\end{array}
\right], \label{eq:FI_matrices}
\end{align}
with
\begin{align}
q &:=\frac{a^{2}r\sin^{2}\theta
          -M\!\left(r^{2}-a^{2}\cos^{2}\theta\right)}
         {\rho^{3}\sqrt{\Delta}},\quad p :=\frac{a^{2}\sin\theta\cos\theta}{\rho^{3}},\quad u :=\frac{ar\sin\theta}{\rho^{3}}, \nonumber \\[0.8ex]
v &:=\frac{a\sqrt{\Delta}\cos\theta}{\rho^{3}}, \quad s:=\frac{r\sqrt{\Delta}}{\rho^{3}}, \quad w :=\frac{(r^{2}+a^{2})\cot\theta}{\rho^{3}}. \label{eq:qpuvsw_defs}
\end{align}

With this compact notation we can now write out the Dirac equation of Eq.~(\ref{eq:arbdireqn})  in the Kerr spacetime as
\begin{align}
\Biggl[
\,i\,\gamma^{\hat{\alpha}}\partial_{\hat{\alpha}}
+i\Bigl(
\Bigl(s-\tfrac12 q\Bigr)\gamma^{\hat1}
+\Bigl(\tfrac12 w-p\Bigr)\gamma^{\hat2}
+\frac{i}{2}v\,\gamma^{\hat1}\gamma^{5}
+\frac{i}{2}u\,\gamma^{\hat2}\gamma^{5}
\Bigr)
-m
\Biggr]\Psi(x)=0\,,
\end{align}
with $\gamma^{5}:=i\,\gamma^{\hat0}\gamma^{\hat1}\gamma^{\hat2}\gamma^{\hat3}$. Explicitly, the four equations are:
\begin{subequations}
\begin{align}
0={}&
 i\,\frac{r^{2}+a^{2}}{\rho\sqrt{\Delta}}\;\partial_{t}\Psi_{1}(x)
+i\,\frac{a}{\rho\sqrt{\Delta}}\;\partial_{\phi}\Psi_{1}(x)
+i\,\frac{a\sin\theta}{\rho}\;\partial_{t}\Psi_{3}(x)
+i\,\frac{1}{\rho\sin\theta}\;\partial_{\phi}\Psi_{3}(x)
+i\,\frac{\sqrt{\Delta}}{\rho}\;\partial_{r}\Psi_{4}(x)
\nonumber\\
&\quad
+\frac{1}{\rho}\;\partial_{\theta}\Psi_{4}(x)
+\frac{i u-v}{2}\,\Psi_{2}(x)
+\frac{w-2p+i(2s-q)}{2}\,\Psi_{4}(x)
-m\,\Psi_{1}(x),
\label{eq:kerrdirac1}
\\[1.0ex]
0={}&
 i\,\frac{r^{2}+a^{2}}{\rho\sqrt{\Delta}}\;\partial_{t}\Psi_{2}(x)
+i\,\frac{a}{\rho\sqrt{\Delta}}\;\partial_{\phi}\Psi_{2}(x)
-i\,\frac{a\sin\theta}{\rho}\;\partial_{t}\Psi_{4}(x)
-i\,\frac{1}{\rho\sin\theta}\;\partial_{\phi}\Psi_{4}(x)
+i\,\frac{\sqrt{\Delta}}{\rho}\;\partial_{r}\Psi_{3}(x)
\nonumber\\
&\quad
-\frac{1}{\rho}\;\partial_{\theta}\Psi_{3}(x)
-\frac{i u+v}{2}\,\Psi_{1}(x)
+\frac{2p-w+i(2s-q)}{2}\,\Psi_{3}(x)
-m\,\Psi_{2}(x),
\label{eq:kerrdirac2}
\\[1.0ex]
0={}&
-i\,\frac{r^{2}+a^{2}}{\rho\sqrt{\Delta}}\;\partial_{t}\Psi_{3}(x)
-i\,\frac{a}{\rho\sqrt{\Delta}}\;\partial_{\phi}\Psi_{3}(x)
-i\,\frac{a\sin\theta}{\rho}\;\partial_{t}\Psi_{1}(x)
-i\,\frac{1}{\rho\sin\theta}\;\partial_{\phi}\Psi_{1}(x)
-i\,\frac{\sqrt{\Delta}}{\rho}\;\partial_{r}\Psi_{2}(x)
\nonumber\\
&\quad
-\frac{1}{\rho}\;\partial_{\theta}\Psi_{2}(x)
+\frac{2p-w+i(q-2s)}{2}\,\Psi_{2}(x)
+\frac{-i u+v}{2}\,\Psi_{4}(x)
-m\,\Psi_{3}(x),
\label{eq:kerrdirac3}
\\[1.0ex]
0={}&
-i\,\frac{r^{2}+a^{2}}{\rho\sqrt{\Delta}}\;\partial_{t}\Psi_{4}(x)
-i\,\frac{a}{\rho\sqrt{\Delta}}\;\partial_{\phi}\Psi_{4}(x)
+i\,\frac{a\sin\theta}{\rho}\;\partial_{t}\Psi_{2}(x)
+i\,\frac{1}{\rho\sin\theta}\;\partial_{\phi}\Psi_{2}(x)
-i\,\frac{\sqrt{\Delta}}{\rho}\;\partial_{r}\Psi_{1}(x)
\nonumber\\
&\quad
+\frac{1}{\rho}\;\partial_{\theta}\Psi_{1}(x)
+\frac{w-2p+i(q-2s)}{2}\,\Psi_{1}(x)
+\frac{i u+v}{2}\,\Psi_{3}(x)
-m\,\Psi_{4}(x).
\label{eq:kerrdirac4}
\end{align}
\end{subequations}


\section{Conclusion \label{sec:conclusion}}
In this manuscript we have presented a pedagogical discussion of the Dirac
equation in arbitrary coordinate systems. Wherever possible, the presentation
is carried out in orthonormal frames, expressed concretely in terms of the
Ricci rotation coefficients, since this is the setting most natural for physical measurements
and closest to the local Lorentz structure of the theory. The discussion also
allows for the use of either constant or position-dependent Dirac matrices
within the orthonormal frame, provided the local Clifford algebra is maintained;
this extra flexibility can be convenient in explicit calculations.

A deliberate emphasis throughout has been on motivation rather than on lengthy
derivations: when a new geometric object is introduced, a simple explanation is given regarding what it is
doing and why it is needed. For example, the Fock-Ivanenko coefficients are
needed so that the spinor covariant derivative is compatible with the Clifford
structure (equivalently, so that the Dirac matrices are covariantly constant),
and the Ricci rotation coefficients encode the corresponding compatibility for
the tetrad/coframe. Along the way a number of practical
consistency conditions, such as adjoint or Hermiticity-type requirements and
related compatibility relations, were also presented that these objects in the Dirac theory
must satisfy. These conditions are useful both conceptually and, importantly, as a diagnostic since
they provide immediate checks for readers who wish to compute the relevant quantities
for their own particular scenario.

The extension to include gravitation implemented as spacetime curvature (with
no torsion) follows in a straightforward manner from the exact same
framework. Finally, we provide several sample calculations in several
important coordinate systems and spacetimes, intended as templates (and checks) for readers who wish to set up the Dirac equation in their own
preferred coordinates.

\section*{Author contributions and declarations}

All authors contributed equally to this work
and have no conflicts of interest to disclose.



\providecommand{\href}[2]{#2}\begingroup\raggedright\endgroup


\appendix

\section{The holonomic or coordinate frame}\label{app:holonomic}
In some advanced settings, such as studies of the Dirac equation coupled to gravitational fields, calculations are often performed in holonomic bases instead of in (the more physical) orthonormal frames. As stated in the introduction, the Dirac equation itself is insensitive to this choice although other physical quantities, such as for example the stress-energy tensor and current, do differ between their holonomic and orthonormal variants.

In the holonomic framework the analogous important expressions can be summarized as follows:

The Dirac equation of Eq.~(\ref{eq:arbdireqn}) can simply be re-written utilizing holonomic indices:
\begin{equation}
\Bigl(i\gamma^{\mu}\nabla_{\mu}-m\Bigr)\Psi(x) = \Bigl[i\gamma^{\mu}\left(\partial_{\mu} + \Gamma_{\mu}(x)\right) -m\Bigr]\Psi(x)  = 0\,, \label{eq:holdireqn}
\end{equation}
with the (possibly spacetime dependent) gamma matrices satisfying
\begin{equation}
 \left\{\gamma^{\mu}(x),\, \gamma^{\nu}(x)\right\}:=\gamma^{\mu}(x)\gamma^{\nu}(x)+\gamma^{\nu}(x)\gamma^{\mu}(x)=2g^{\mu\nu}(x) \mathbbm{1}\,. \label{eq:holonclifalg}
\end{equation}

The connection $\Gamma_{\mu}$ is often known as the \emph{spinor affine connection},
\begin{equation}
\Gamma_{\mu}(x)
=
\frac{1}{4}\,\gamma_{\hat{\alpha}}(x)\!
\left(
\partial_{\mu}\gamma^{\hat{\alpha}}(x)
+
\omega_{\mu}{}^{\hat{\alpha}}{}_{\hat{\beta}}(x)\,
\gamma^{\hat{\beta}}(x)
\right)+\; a_{\mu}(x)\,\mathbbm{1}
\,, \label{eq:spinaffinecon}
\end{equation}
with $a_{\mu}(x)$ the coordinate gradient of an imaginary function, usually taken as zero. The spinor affine connection now possesses the 
Hermiticity property
\begin{equation}
\Gamma_{\mu}^{\dagger}(x)
=
-\gamma^{\hat{0}}(x)\,\Gamma_{\mu}(x)\,\gamma^{\hat{0}}(x)
+
\big(\partial_{\mu}\gamma^{\hat{0}}(x)\big)\,\gamma^{\hat{0}}(x)\, ,
\label{eq:hermcond_Gammamu}
\end{equation}
and
the covariant derivative of the holonomic gamma matrices, analogous to Eq.~(\ref{eq:covdifgamma}) is
\begin{equation}
\nabla_{\mu}\gamma^{\nu}(x)\;:=\;
\partial_{\mu}\gamma^{\nu}(x)
+\Gamma^{\nu}{}_{\mu\rho}(x)\,\gamma^{\rho}(x)
+\big[\Gamma_{\mu}(x),\,\gamma^{\nu}(x)\big]
\equiv 0 \,.
\label{eq:covdifgamma_hol}
\end{equation}

The new connection introduced above, $\Gamma^{\nu}{}_{\mu\rho}(x)$, is the \emph{Christoffel (or
Levi-Civita) connection}, whose coordinate components are the \emph{Christoffel symbols}:
\begin{equation}
\Gamma^{\nu}{}_{\mu\rho}(x)
:=
\frac{1}{2}\,g^{\nu\sigma}(x)
\Bigl[
\partial_{\mu}g_{\sigma\rho}(x)
+\partial_{\rho}g_{\sigma\mu}(x)
-\partial_{\sigma}g_{\mu\rho}(x)
\Bigr]\, . \label{eq:chris}
\end{equation}

In terms of the Christoffel symbols the spinor affine connection can also be written as
\begin{equation}
\Gamma_{\mu}(x)
=
\frac{1}{4}\,\gamma_{\nu}(x)
\Bigl(
\partial_{\mu}\gamma^{\nu}(x)
+
\Gamma^{\nu}{}_{\mu\rho}(x)\,\gamma^{\rho}(x)
\Bigr)
\;+\; a_{\mu}(x)\,\mathbbm{1}\, ,
\label{eq:spinaffinecon_chris}
\end{equation}
where as before $a_{\mu}(x)$ is usually set to zero.

We also mention here that in holonomic components the analog of Eq.~(\ref{eq:orthgamherm}) is
\begin{equation}
 \left(\gamma^{\mu}\right)^{\dagger}=\gamma^{\hat{0}}\gamma^{\mu}\gamma^{\hat{0}}\, \label{eq:coordgamherm}
\end{equation} 
(note that \emph{orthonormal} $\gamma^{\hat{0}}$ still appears here). However it is generally
\textbf{not} true that $(\gamma^{0})^{\dagger}=\gamma^{0}$ and
$(\gamma^{j})^{\dagger}=-\gamma^{j}$ for holonomic $\gamma^{\mu}$.

Finally, we mention that the Dirac adjoint is still given by
$\overline{\Psi}(x):=\Psi^\dagger(x)\,\gamma^{\hat{0}}(x)$ \footnote{It should be noted that even if one is using holonomic bases $\overline\Psi(x)$ is constructed out of the \emph{orthonormal} $\gamma^{\hat{0}}$. See \protect\cite{ref:yaupsibar}, \protect\cite{ref:collasandklein}.}.


\section{A method to calculate the tetrad}\label{app:tetrad}
When one actually wants to construct a tetrad explicitly in a given coordinate
system, a practical approach is to begin with a convenient set of four linearly
independent covectors (1-forms) written in holonomic components, say
\begin{equation}
v^{(0)}_{\mu}(x),\quad v^{(1)}_{\mu}(x),\quad v^{(2)}_{\mu}(x),\quad
v^{(3)}_{\mu}(x),
\end{equation}
and then orthonormalize them using the metric $g_{\mu\nu}(x)$ as the inner
product (the holonomic basis metric is used since the $\mu$ index is a holonomic one), exactly as in the Gram--Schmidt procedure from linear algebra. We ignore here the special case where an intermediate Gram–Schmidt step produces a null (zero-norm) candidate. One
builds the orthonormal coframe vectors $e_{\mu}{}^{\hat{\alpha}}(x)$ step by step.
First choose a timelike direction and normalize it. That is, one chooses any $v^{(0)}_{\mu}(x)$ that respects the condition $v^{(0)}_{\mu}(x)v^{(0)}_{\nu}(x)g^{\mu\nu}(x)> 0$ in the mostly-minus
convention. The normalization condition is then
\begin{equation}
g^{\mu\nu}\,e_{\mu}{}^{\hat{0}}\,e_{\nu}{}^{\hat{0}}
=\eta^{\hat{0}\hat{0}}=+1,
\end{equation}
so to achieve this from our chosen $v^{(0)}_{\mu}(x)$ we set
\begin{equation}
e_{\mu}{}^{\hat{0}}
=
\frac{v^{(0)}_{\mu}}
{\sqrt{\left|\,v^{(0)}_{\rho}\,g^{\rho\sigma}\,v^{(0)}_{\sigma}\,\right|}}\,.
\end{equation}
Next take a candidate spacelike covector $v^{(1)}_{\mu}$ (i.e. one not collinear with $v^{(0)}_{\mu}(x)$), remove its component
along $e_{\mu}{}^{\hat{0}}$ by subtracting the projection, and then normalize.
The projection coefficient is
\begin{equation}
g^{\rho\sigma}\,v^{(1)}_{\rho}\,e_{\sigma}{}^{\hat{0}},
\end{equation}
so the orthogonalized covector is
\begin{equation}
\big(v^{(1)}_{\perp}\big)_{\mu}
=
v^{(1)}_{\mu}
-
\big(g^{\rho\sigma}\,v^{(1)}_{\rho}\,e_{\sigma}{}^{\hat{0}}\big)\,
e_{\mu}{}^{\hat{0}}.
\end{equation}
Then normalize it so that
\begin{equation}
g^{\mu\nu}\,e_{\mu}{}^{\hat{1}}\,e_{\nu}{}^{\hat{1}}
=\eta^{\hat{1}\hat{1}}=-1,
\end{equation}
namely
\begin{equation}
e_{\mu}{}^{\hat{1}}
=
\frac{\big(v^{(1)}_{\perp}\big)_{\mu}}
{\sqrt{\left|\,\big(v^{(1)}_{\perp}\big)_{\rho}\,g^{\rho\sigma}\,
\big(v^{(1)}_{\perp}\big)_{\sigma}\,\right|}}.
\end{equation}
One repeats the same idea for the remaining directions. For example, starting
from $v^{(2)}_{\mu}$ one subtracts projections onto both $e_{\mu}{}^{\hat{0}}$
and $e_{\mu}{}^{\hat{1}}$:
\begin{equation}
\big(v^{(2)}_{\perp}\big)_{\mu}
=
v^{(2)}_{\mu}
-
\big(g^{\rho\sigma}\,v^{(2)}_{\rho}\,e_{\sigma}{}^{\hat{0}}\big)\,
e_{\mu}{}^{\hat{0}}
-
\big(g^{\rho\sigma}\,v^{(2)}_{\rho}\,e_{\sigma}{}^{\hat{1}}\big)\,
e_{\mu}{}^{\hat{1}},
\end{equation}
and then normalizes to obtain $e_{\mu}{}^{\hat{2}}$ with
\begin{equation}
g^{\mu\nu}\,e_{\mu}{}^{\hat{2}}\,e_{\nu}{}^{\hat{2}}
=\eta^{\hat{2}\hat{2}}=-1.
\end{equation}
Likewise for $e_{\mu}{}^{\hat{3}}$, subtracting projections onto
$e_{\mu}{}^{\hat{0}},\,e_{\mu}{}^{\hat{1}},\,e_{\mu}{}^{\hat{2}}$ and then
normalizing. At the end of the procedure one has
\begin{equation}
g^{\mu\nu}\,e_{\mu}{}^{\hat{\alpha}}\,e_{\nu}{}^{\hat{\beta}}
=
\eta^{\hat{\alpha}\hat{\beta}}.
\end{equation}

This procedure may seem a bit complicated due to its generality, but often it is quite straightforward to compute a suitable tetrad without employing all of this machinery. This is explicitly illustrated in section \ref{subsec:spherical}.

Different reasonable starting choices $v^{(a)}_{\mu}$, or different ways of
ordering the Gram-Schmidt steps, lead to different tetrads.


\end{document}